\documentclass[taps,amsmath,amssymb,floatfix, twocolumn, pre
]{revtex4-2}

\usepackage{graphicx}% Include figure files
\usepackage{dcolumn}% Align table columns on decimal point
\usepackage{bm}% bold math
\graphicspath{{./pic/}}
\usepackage{color}
\usepackage{amsmath}
\usepackage{tabularx}
\usepackage{rotating}
\usepackage{epic}
\usepackage{subfigure}
\usepackage{caption}
\usepackage{hyperref}
\begin{document}

\preprint{APS/123-QED}

\title{Thermal capillary waves on bounded nanoscale thin films  }% Force line breaks with \\

\author{Jingbang Liu}
\email{jingbang.liu@warwick.ac.uk}
\affiliation{
Mathematics Institute, University of Warwick, Coventry CV4 7AL, United Kingdom
}

\author{Chengxi Zhao}
\email{zhaochengxi@ustc.edu.cn}
\affiliation{Department of Modern Mechanics, University of Science and Technology of China, Hefei 230026, China}

\author{Duncan A. Lockerby}
 \email{D.Lockerby@warwick.ac.uk}
\affiliation{School of Engineering, University of Warwick, Coventry CV4 7AL, United Kingdom }

\author{James E. Sprittles}
\email{J.E.Sprittles@warwick.ac.uk}
\affiliation{Mathematics Institute, University of Warwick, Coventry CV4 7AL, United Kingdom
}

\begin{abstract}
    The effect of confining walls on the fluctuation of a nanoscale thin film's free surface is studied using the stochastic thin-film equations (STFE). Two canonical boundary conditions are employed to reveal the influence of the confinement: (i) an imposed contact angle and (ii) a pinned contact line. A linear stability analysis provides the wave eigenmodes, after which thermal-capillary-wave theory predicts the wave fluctuation amplitudes. Molecular dynamics (MD) simulations are performed to test the predictions and a Langevin diffusion model is proposed to capture oscillations of the contact-lines observed in MD. Good agreement between the theoretical predictions and the MD simulation results is recovered, and it is discovered that confinement can influence the entire film. Notably, a constraint on the length scale of wave modes is found to affect fluctuation amplitudes from our theoretical model, especially for 3D films. This opens up new challenges and future lines of inquiry.

\end{abstract}
\maketitle

\section{Introduction}

The behavior of fluids at the nanoscale attracts increasing attention as fluid-based technologies continue to minituarize \cite{bocquetNanofluidicsBulkInterfaces2010}, for example, in: lab-on-a-chip devices \cite{craigheadFutureLabonachipTechnologies2006}, nanofluidic transistors \cite{karnikElectrostaticControlIons2005a}, ink-jet printing \cite{basaranNonstandardInkjets2013} and osmotic transport \cite{lokeshOsmoticTransportSurface2018}. The dynamics at such scales are challenging, if not impossible, to observe experimentally, making modeling and simulation a vital component of continued technological progress. However, due to the additional physical phenomena that appear when going from traditional engineering scales to the nanoscale \cite{kavokineFluidsNanoscaleContinuum2021}, conventional fluid dynamical modeling approaches are often inaccurate.

A canonical nanoscale flow topic that underpins many applications is the behavior and stability of thin liquid films on rigid solid surfaces. Here, stability is crucial to coating technologies \cite{weinsteinCOATINGFLOWS2004,kumarLiquidTransferPrinting2015} whilst instability can be harnessed to create pre-determined patterns \cite{crasterDynamicsStabilityThin2009}. Driven by technological demands and fundamental interest, there is a huge body of research in this field, see for example review articles \cite{crasterDynamicsStabilityThin2009,oronLongscaleEvolutionThin1997,myersThinFilmsHigh1998,bonnWettingSpreading2009}.  It is well established that at the nanoscale disjoining pressure becomes important, competing with surface tension for the stability of the film and driving rupture; via the so-called spinodal mechanism \cite{beckerComplexDewettingScenarios2003,sharmaManyPathsDewetting2003,alizadehpahlavanThinFilmsPartial2018}.  Notably, though, in order for theoretical predictions of rupture timescales to agree with those from experiment, thermal fluctuations, which drive free-surface nanowaves, need to be incorporated in the physical model  \cite{fetzerThermalNoiseInfluences2007}. The dynamics of these nanowaves on thin films form the basis of this work, where we consider, for the first time, their behavior within a confined environment, i.e. bounded by surfaces.

It has long been expected that the chaotic thermal motion of molecules in a liquid would generate so-called `thermal capillary waves' at liquid-fluid interfaces \cite{mandelstamUeberRauhigkeitFreier1913,bouchiatSpectreFluctuationsThermiques1971,vinkCapillaryWavesColloidpolymer2005}. One of the earliest experimental confirmations of the existence of such waves was obtained using a light-scattering technique at the liquid-vapour interface of carbon dioxide \cite{bouchiatLightScatteringSurface1969}. More recently, experiments have been conducted to observe and measure thermal capillary waves by exploiting ultra-low surface tension fluids that generate micron-scale waves \cite{aartsDirectVisualObservation2004,derksSuppressionThermallyExcited2006}, using various optical scattering techniques \cite{liCoupledCapillaryWave2001,huObservationLowviscosityInterface2006,alvineCapillaryWaveDynamics2012}, and, with simple fluids, using an atomic force microscope cantilever placed on a micro hemispherical bubble \cite{zhangNearFieldProbeThermal2021}. 

An alternative tool for probing the physics of the nanoscale is molecular dynamics (MD) simulations, providing an environment for conducting `virtual experiments'  \cite{koplikMolecularDynamicsPoiseuille1988,delgado-buscalioniHydrodynamicsNanoscopicCapillary2008} that complement traditional methods and yield additional understanding. MD simulations have observed thermal capillary waves in the context of: nanoscale thin films \cite{zhangMolecularSimulationThin2019,zhangRelaxationThermalCapillary2021,zhangThermalCapillaryWave2021,thakreFiniteSystemSize2008,willisThermalCapillaryWaves2010}, the instability and breakup of liquid jets \cite{moselerFormationStabilityBreakup2000,zhaoRevisitingRayleighPlateau2019,zhaoDynamicsLiquidNanothreads2020}, the coalescence of nanodroplets \cite{perumanathDropletCoalescenceInitiated2019,perumanathMolecularPhysicsJumping2020,pothierMoleculardynamicsStudyViscous2012} and  films on fibers \cite{zhangNanoscaleThinfilmFlows2020}. Whilst MD contains the necessary nanoscale physics to capture thermal capillary waves, it is both very computationally expensive and requires interpretation that is arguably best provided by macroscopic theories. For illustration, in this article, a $51.4$ ns simulation of a thin film containing $32883$ Lennard-Jones particles took $11$ hours to run on a $28$ core CPU; and to obtain statistically reliable averages, multiple realizations are needed. Clearly, there is a need for a complementary modeling approach that is more computationally tractable. 

To go beyond conventional fluid mechanics and include thermal fluctuations, Landau and Lifshitz  introduced the equations of fluctuating hydrodynamics (FH) \cite{landauFluidMechanics1995} by adding a random stress tensor satisfying the fluctuation dissipation theorem into the Navier-Stokes equations. For thin liquid films, the stochastic thin-film equation (STFE), accurate in the lubrication approximation, has been derived for planar films \cite{davidovitchSpreadingViscousFluid2005,grunThinFilmFlowInfluenced2006}; a similar stochastic equation has been obtained for jets \cite{moselerFormationStabilityBreakup2000}. Extensions of the STFE have also been derived, for example, for different slip conditions \cite{zhangNanoscaleThinfilmFlows2020,zhangRelaxationThermalCapillary2021} and with an elastic plate on top of the film \cite{pedersenAsymptoticRegimesElastohydrodynamic2019}.

A linear stability analysis can be applied to the STFE to obtain a power spectrum for the thermal capillary waves \cite{meckeThermalFluctuationsThin2005} that can be compared with experiment \cite{fetzerThermalNoiseInfluences2007}. The power spectrum of the free-surface waves has also been shown to agree with MD \cite{zhangMolecularSimulationThin2019,zhangNanoscaleThinfilmFlows2020}, exhibiting unconventional effects like an evolving wave number associated with fastest growth. Attempts have also been made to solve the full nonlinear STFE, and its variants, numerically \cite{zhaoFluctuationdrivenDynamicsNanoscale2022,zhaoDynamicsLiquidNanothreads2020,shahThermalFluctuationsCapillary2019,duran-olivenciaInstabilityRuptureFluctuations2019} which, although requiring more complex formulations, still generate large computational savings compared with MD. In summary, the STFE is a remarkably powerful and efficient tool for studying the dynamics of ultra-thin films whose potential is yet to be fully exploited (e.g., thus far most analyses are confined to 2D).

Notably, previous studies of the STFE either assume the films are unbounded, that the dynamics are periodic on some length scale (essentially, to enable a simple Fourier analysis), or that the boundaries are sufficiently far away that they have no effect other than to potentially regularize the solution at some upper scale. How then, does confinement, i.e. the effects of nearby boundaries, affect the dynamics of nanoscale films? This will be our focus, beginning by considering the properties of nanowaves in thermal equilibrium.

In this work, we examine, in both quasi-2D and 3D, the effect of the two typical boundary conditions where a free surface meets a wall: (i) an imposed contact angle and (ii) a (partially) pinned contact line. A linear stability analysis is performed and the waves modes are calculated by solving the eigenvalue problem for each boundary condition. Thermal-capillary-wave theory is used to predict the fluctuation amplitude and then validated against MD simulations.

The paper is organized as follows. In Section~\ref{sec:2d} we consider quasi-2D bounded films with the two different boundary conditions and for each provide two ways to derive a theoretical prediction for the fluctuation amplitude of the free surface: (i) from thermal-capillary-wave theory; and (ii) directly from the STFE. Details of MD simulations are provided and results are compared with the theory. In Section~\ref{sec:3d} we extend our study to 3D circular bounded films with two different boundary conditions; theories and fluctuation amplitudes are derived. MD simulations are performed and results are compared. In Section~\ref{sec:cutoff} the role of a cut-off length scale is then discussed. In Section~\ref{app:future}, future research directions are outlined.

\section{Quasi-2D bounded thin films}
\label{sec:2d}

\begin{figure}
    \centering
    \includegraphics[width=\linewidth]{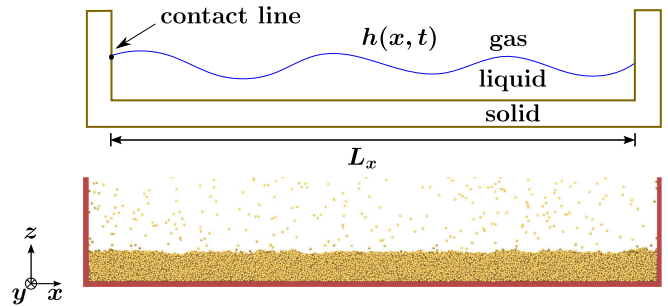}
    \caption{An illustration of the geometry of the quasi-2D thin-film problem (top) and a snapshot of a representative MD simulation for a thin film with $90^\circ$ contact angle (below); yellow particles denote liquid Argon and red particles denote Platinum solid.}
    \label{fig:thinfilm}
\end{figure}

In this section, we present the modeling and MD simulation results of 2D bounded thin films on a solid that is in the $(x,y)$-plane at $z=0$, as shown in Fig.~\ref{fig:thinfilm}. The MD simulations are inherently 3D, and to approximate a 2D flow the thickness of the film $L_y$ in the $y$-direction is set to be much smaller than the length of the film $L_x$, making it `quasi-2D'. To compare the theory to MD results, we consider quantities which are averaged `into the page', over $L_y$ in the $y$-direction, resulting in all quantities depending only on $(x,t)$, see \cite{zhangMolecularSimulationThin2019}. Assuming that $(H/L_x)^2Re\ll 1$, where $H$ is the characteristic height of the free surface and $Re$ is the Reynolds number, which is expected to be true for nanoscale thin films at thermal equilibrium, we can apply the lubrication approximation \cite{oronLongscaleEvolutionThin1997} to the Navier-Stokes equations and find that inertial effects are negligible. Then, in the absence of disjoining pressure, whose influence we also assume to be negligible in thermal equilibrium for the film heights we consider, we arrive at the thin-film equation (TFE) to provide a description of the free surface  $z=h(x,t)$ given by
\begin{equation}
    \label{eq:2dLE}
    \frac{\partial h}{\partial t} = -\frac{\gamma}{3\mu}\frac{\partial}{\partial x}\left(h^3\frac{\partial^3h}{\partial x^3}\right),
\end{equation}
where $\gamma$ is the surface tension and $\mu$ is the dynamic viscosity. When thermal fluctuations are included, the stochastic thin-film equation (STFE) \cite{grunThinFilmFlowInfluenced2006,davidovitchSpreadingViscousFluid2005,meckeThermalFluctuationsThin2005} can be derived from fluctuating hydrodynamics:
\begin{equation}
    \label{eq:2dSLE}
    \frac{\partial h}{\partial t} = -\frac{\gamma}{3\mu}\frac{\partial}{\partial x}\left(h^3\frac{\partial^3 h}{\partial x^3}\right)+\sqrt{\frac{2k_BT}{3\mu L_y}}\frac{\partial}{\partial x}(h^{3/2}\mathcal{N}),
\end{equation}
where $k_B$ is the Boltzmann constant, $T$ is the temperature. Thermal noise $\mathcal{N}(x,t)$ has zero mean and covariance
\begin{equation}
\label{eq:2dnoise_property}
    \langle \mathcal{N}(x,t)\mathcal{N}(x',t')\rangle = \delta(x-x')\delta(t-t'),
\end{equation}
which means that the noise is uncorrelated in both time and space.
Note, the $\sqrt{1/L_y}$ factor in the noise term of Eq.~(\ref{eq:2dSLE}) comes from averaging in the $y$-direction. 

One can easily see that a flat free surface $h(x,t)=h_0$ is a steady solution to Eq.~(\ref{eq:2dLE}). However, thermal fluctuations, modeled by the noise term in Eq.~(\ref{eq:2dSLE}), drives the free surface away from the steady solution, creating thermal capillary waves \cite{meckeThermalFluctuationsThin2005,vinkCapillaryWavesColloidpolymer2005,delgado-buscalioniHydrodynamicsNanoscopicCapillary2008,crasterDynamicsStabilityThin2009,willisThermalCapillaryWaves2010}. Note, these `waves' are viscous damped response to fluctuations arising from within the bulk liquid of the film (i.e. they are inertia free). In the case of fluctuation-driven films, $\langle h(x,t) \rangle=h_0$, where $\langle \rangle$ represents ensemble average.   

To understand the properties of the nanoscale waves, we consider a linearized setup with $h(x,t)=h_0+\delta h(x,t)$ and $\delta h\ll h_0$. Then, as is conventionally done, if we assume the domain is periodic on a length $L_x$, the perturbation can be decomposed into Fourier modes and the fluctuation amplitude (or surface roughness) can be estimated by \cite{vinkCapillaryWavesColloidpolymer2005,delgado-buscalioniHydrodynamicsNanoscopicCapillary2008}
\begin{equation}
\label{eq:TCW_periodic}
    \langle \delta h^2\rangle = \frac{l_T^2}{12}\frac{L_x}{L_y},
\end{equation}
where $l_T=\sqrt{k_B T/\gamma}$ is the `thermal length scale' characterizing the approximate amplitude of these waves. Interestingly, the dynamic growth of these nanoscale waves from an initially flat interface has been shown in \cite{zhangThermalCapillaryWave2021} to fall into a specific universality class.

Here, we consider a different, practically more realistic setup, with solid walls at $x=0$, $L_x$ and two different physically inspired boundary conditions: (i) a prescribed $90^\circ$ contact angle; and (ii) (partially) pinned contact lines. 

\subsection{Prescribed $90^\circ$ contact angle}

As a starting point, we consider a $90^\circ$ contact angle, for which the equilibrium state is on average a flat film. In this case
\begin{equation}
    \label{eq:BC90}
	\left.\frac{\partial h}{\partial x}\right|_{x=0}=\left.\frac{\partial h}{\partial x}\right|_{x=L_x}=0.
\end{equation}
It is worth noting that here we have assumed that the contact angle is $90^\circ$ at every instant in time. Assuming also that the walls are impermeable, we have
\begin{equation}
    \label{eq:BCnoflux}
	\left.\frac{\partial^3 h}{\partial x^3}\right|_{x=0} = \left.\frac{\partial^3 h}{\partial x^3}\right|_{x=L_x} = 0.
\end{equation}
Since the boundary conditions are not periodic, we can no longer assume the wave modes to be Fourier. Instead, linearizing the TFE and solving the corresponding eigenvalue problem, we can show that the appropriate wave modes (see Appendix \ref{app:modes_90}) are as follows:
\begin{equation}
    \phi_n(x) = \cos\left(\frac{n\pi x}{L_x}\right),\: n=1,2,\ldots
\end{equation}
Given this information, we can proceed with the classical `thermal-capillary-wave theory' approach \cite{grantFluctuatingHydrodynamicsCapillary1983}.

\subsubsection{Thermal-capillary-wave theory}
The free surface can be written as the superposition of the average film thickness $h_0$ and a perturbation $h_1(x,t)$:
\begin{equation}
\label{eq:90_dcomp}
    h(x,t) = h_0 + h_1(x,t).
\end{equation}
Here, $h_1(x,t)$ can be decomposed into the wave modes $\phi_n(x)$, so that
\begin{equation}
    \label{eq:90_h1}
    h_1(x,t) = \sum_{n=1}^{\infty}a_n(t)\phi_n(x),
\end{equation}
and it is assumed that $a_n(t)\ll h_0$. An energetic argument, exploiting equipartition in thermal equilibrium, will then give us the statistical properties of the amplitudes (the $a_n$'s).

The cost of energy for doing work against surface tension by expanding the interface's area is given by
\begin{equation}
    E = \gamma \left(L_y\int_0^{L_x}\sqrt{1+\left(\frac{\partial h}{\partial x}\right)^2} dx - L_xL_y\right),
\end{equation}
where $L_y$ is the film length into the page. Taking the standard thin-film approximation that $\partial h/\partial x \ll 1$ we have
\begin{equation}
\label{eq:TCW}
    E \approx \gamma L_y\int_0^{L}\frac{1}{2}\left(\frac{\partial h}{\partial x}\right)^2 dx,
\end{equation}
so that using Eq.~(\ref{eq:90_dcomp}) and Eq.~(\ref{eq:90_h1}) we can obtain the total energy:
\begin{equation}
    E = \sum_{n=1}^{\infty}E_n = \sum_{n=1}^{\infty} \frac{\gamma\pi^2n^2}{4}\frac{L_y}{L_x} a_n^2.
\end{equation}
According to the equipartition theorem, at thermal equilibrium the energy is shared equally among each mode, i.e. $\langle E_n \rangle = k_B T/2$, leading to an expression for the variance of each mode's amplitude (note their means are zero by construction):
\begin{equation}
    \label{eq:equipartition90}
    \langle a_n^2 \rangle = \frac{2}{\pi^2}l_T^2\frac{L_x}{L_y}\frac{1}{n^2}.
\end{equation}

This expression then allows us to obtain information about the nanowaves in thermal equilibrium. Using Eq.~(\ref{eq:90_h1}) and  $\langle a_m a_n\rangle = \delta_{mn}\langle a_n^2\rangle$ (see Appendix \ref{app:90_dh2_SLE}), we can find the variance of the perturbation across the film as follows
\begin{align}
    \notag
    \langle h_1^2(x)\rangle &= \left\langle \sum_{m=1}^{\infty}a_m\cos\left(\frac{m\pi x}{L_x}\right)\sum_{n=1}^{\infty}a_n\cos\left(\frac{n\pi x}{L_x}\right) \right\rangle\\
    \notag
    &=\sum_{n=1}^{\infty}\langle a_n^2\rangle\cos^2\left(\frac{n\pi x}{L_x}\right)\\
    \notag
    &= \frac{2l_T^2}{\pi^2}\frac{L_x}{L_y}\sum_{n=1}^{\infty}\frac{\cos^2\left(\frac{n\pi x}{L_x}\right)}{n^2}\\
    \label{eq:TCW_90}
    &=l_T^2\frac{L_x}{L_y}\left[\frac{1}{12}+\left(\frac{1}{2}-\frac{x}{L_x}\right)^2\right].
\end{align}
Notably, in contrast to the periodic spatially homogeneous case Eq.~(\ref{eq:TCW_periodic}), expression Eq.~(\ref{eq:TCW_90}) is a function of $x$. A full discussion of this case will be provided after we have compared to MD results. 

There is also an alternative derivation for $\langle a_n^2\rangle$ directly from the STFE (see Appendix~\ref{app:90_dh2_SLE}). Since the STFE describes the time evolution of the film height from some initial (non-equilibrium) state, the result is also time dependent:
\begin{equation}
    \langle a_n^2\rangle = \frac{2k_B T}{\gamma \pi^2}\frac{L_x}{L_y}\frac{1}{n^2}(1-\exp(-2An^4t)),
\end{equation}
where $A=\gamma h_0^3 \pi^4/(3\mu L_x^4)$. This tells us that an initial perturbation decays exponentially with time, and that at thermal equilibrium (as $t\to\infty$) the results from the STFE agrees with Eq.~(\ref{eq:equipartition90}) derived from thermal-capillary-wave theory, which provides a more straight forward derivation.

\subsubsection{Molecular-dynamics simulations}
\label{sec:2d_90_MD}
\begin{table}[b]
\caption{\label{tab:90}Simulation parameters and their non-dimensional values (reduced units based on Lennard-Jones potential parameters $\epsilon_{f\!f}$, $\sigma_{f\!f}$, $m_f$) for a $90^\circ$ contact angle.
}
\begin{ruledtabular}
\begin{tabular}{cccc}
Property &Nondim.value & Value & Unit  \\
\hline\\[-2ex]
$\epsilon_{f\!f}$ & 1 & $1.67\times 10^{-21}$ & J \\
$\epsilon_{sf}$ & 0.52 & $0.8684\times 10^{-21}$ & J \\
$\epsilon_{ss}$ & 50 & $83.5\times 10^{-21}$ & J \\
$\sigma_{f\!f}$ & 1 & $0.34$ & nm \\
$\sigma_{sf}$ & 0.8 & $0.272$ & nm \\
$\sigma_{ss}$ & 0.72 & $0.247$ & nm  \\
$m_f$ & 1 & $6.63\times 10^{-26}$ & kg  \\
$m_s$ & 4.8863 & $32.4\times 10^{-26}$ & kg \\
$T$ & 0.7 & $85$ & $K$ \\
$\rho_l$ & 0.83 & $1.4\times 10^3$ & kg/m$^3$  \\
$\rho_v$ & 0.0025 & $3.5$ & kg/m$^3$  \\
$\rho_s$ & 2.6 & $21.45 \times 10^3$ & kg/m$^3$  \\
$r_c$ & 5.5 & $1.87$ & nm 
\end{tabular}
\end{ruledtabular}
\end{table}

To verify our new theoretical prediction, we use molecular dynamics simulations (MD) as a virtual experiment to probe the behavior of quasi-2D thin films that are bounded on both sides by solid walls with $90^\circ$ contact angles.

The simulations are performed in the open-source software LAMMPS \cite{thompsonLAMMPSFlexibleSimulation2022}, which has been widely used to study fluid phenomena at the nanoscale, e.g. \cite{zhangMolecularSimulationThin2019,zhaoRevisitingRayleighPlateau2019,fernandez-toledanoContactlineFluctuationsDynamic2019,fernandez-toledanoMoleculardynamicsStudySliding2019,chaconEffectDispersionForces2014,nguyenRuptureMechanismLiquid2014,wangMolecularOriginContact2015,wenMolecularDropletsVs2021,theodorakisDropletControlBased2021,nguyenCoexistenceSpinodalInstability2014}

Argon is used as a fluid and Platinum is used for the solid walls. The interaction between particles are modeled using the conventional Lennard-Jones 12-6 potential
\begin{equation}
    V(r_{ij}) = 4\epsilon_{AB}\left[\left(\frac{\sigma_{AB}}{r_{ij}}\right)^{12}-\left(\frac{\sigma_{AB}}{r_{ij}}\right)^{6}\right],
\end{equation}
where $r_{ij}$ is the distance between atoms $i$ and $j$, $\epsilon_{AB}$ is the energy parameter representing the depth of potential wells and $\sigma_{AB}$ is the length parameter representing the effective atomic diameter. Here, $AB$ are different combinations of particle types; namely, fluid-fluid (ff), solid-solid (ss) and solid-fluid (sf). The simulation parameters are summarized in Table~\ref{tab:90} with corresponding non-dimensional `MD values' henceforth denoted with an asterisk (as one can see, energy is scaled with respect to $\epsilon_{f\!f}$, lengths with $\sigma_{f\!f}$ and mass with $m_f$). To obtain a $90^\circ$ contact angle, we set $\epsilon^*_{sf}=0.52$ and $\sigma_{sf}^{*}=0.8$. The position of solid particles are fixed to reduce computational cost. The timestep is set to $0.0085$ ps.

Transport properties of liquid Argon are measured under MD simulations, with parameters given by Table~\ref{tab:90}. Shear viscosity $\mu=2.44\times 10^{-4}$ kg/(ms) is calculated using the Green-Kubo method \cite{haileMolecularDynamicsSimulation1993}:
\begin{equation}
    \mu = \frac{V}{3k_B T}\sum\int_0^{\infty}\langle J_{pq}(t)J_{pq}(0) \rangle dt,
\end{equation}
where $V$ is the volume, $J_{pq}$ are the components of the stress tensor and the sum accumulates three terms given by $pq$ ($=xy,yz,zx$). Note, only off-diagonal terms of the stress tensor are used, as shear viscosity is measured by the transport of momentum perpendicular to velocity. Surface tension $\gamma=1.52\times 10^{-2}$ N/m is calculated from the difference between the normal and tangential components of pressure tensor in a simple vapor-liquid-vapor system ($z$-direction) \cite{trokhymchukComputerSimulationsLiquid1999,smithLangevinModelFluctuating2016}:
\begin{equation}
    \gamma = \frac{1}{2}\int_0^{L_z}\left(P_{zz}(z)-\frac{1}{2}\left(P_{xx}(z)+P_{yy}(z)\right)\right)dz,
\end{equation}
where $L_z$ is the length of the simulation box in the $z$-direction and $P_{xx}$, $P_{yy}$, $P_{zz}$, are the diagonal components of the pressure tensor.

To set up the MD simulation we use the following procedure: (i) a block of liquid Argon is created in a periodic box with dimension $(L_x,L_y,h_0)$, density $\rho_l$  and is equilibrated for $5\times 10^6$ timesteps with NVT at temperature $T$, (ii) a block of vapor Argon is created in a periodic box with dimension $(L_x,L_y,3h_0)$, density $\rho_v$ then equilibrated for $5\times 10^6$ timesteps with NVT at temperature $T$, (iii) Platinum walls are created with a face centered cubic structure (fcc) of density $\rho_s$, each wall has $5$ layers of Platinum atoms with thickness $0.872$ nm, the bottom wall then has dimension $(L_x,L_y,0.872)$ , (iv) the equilibrated liquid Argon is then place onto the bottom wall with a $0.17$ nm gap between the solid and the liquid (the gap results from the repulsive force in the Lennard-Jones potential and its thickness is found after equilibration) \cite{zhangNanoscaleThinfilmFlows2020}, (v)  equilibrated vapor Argon is placed on the top. The MD simulation is then run with NVT at temperature $T$. Fig.~\ref{fig:thinfilm} shows a snapshot of the MD simulation. Periodic boundary conditions are applied only in the $y$-direction. A reflective wall is applied at the top boundary.

In our simulations, the position of the liquid-vapor interface is determined using the number density and a binning technique see \cite{zhangThermalCapillaryWave2021}. We first calculate the number density of each Argon particle using a cut-off radius of $3.5\sigma_{f\!f}$. Particles with number density above $0.5 n^*$ are then defined as liquid particles and particles with number density below $0.5n^*$ are identified as vapor particles, where $n^*=0.83$ is the non-dimensional number density of a liquid Argon particle in the bulk. The simulation domain is uniformly divided into vertical bins and the position of the free surface in each bin is determined by taking the maximum of the heights of all liquid particles inside the bin. Here, we use bins with side length $1.5\sigma_{f\!f}$ in $x$ and $1.4\sigma_{f\!f}$ in $y$.  As a result the free surface position is projected onto a $x-y$ mesh and expressed as a 2D array.

Three different film lengths are tested: Film 1 ($L_x=13.04$ nm), Film 2 ($L_x=25.99$ nm) and Film 3 ($L_x=51.29$ nm). The film width $L_y=2.94$ nm is chosen so that the MD simulation can be consider quasi-2D. The initial film height $h_0=4.85$ nm is chosen so that the film is relatively thin, but yet does not breakup due to disjoining pressure \cite{oronLongscaleEvolutionThin1997,crasterDynamicsStabilityThin2009,zhangMolecularSimulationThin2019}. The equilibriation time $t_e$, i.e. the time taken for all the waves to fully develop from an initially flat interface, is estimated by Eq.~(\ref{eq:app90equT}), which is the characteristic time for the mode with the longest wavelength (and thus slowest growth) to develop; this varies with film length $L_x$. Multiple independent MD simulations (realizations) (Film 1: 10, Film 2: 10, Film 3: 20) are performed in parallel to reduce wall-clock simulation time. Data is gathered after $t_e$ every $4000$ timesteps and the free surface position is averaged in the $y$-direction, to provide $h=h(x,t)$ at each snapshot.

\begin{figure}[h]
\includegraphics[width=\linewidth]{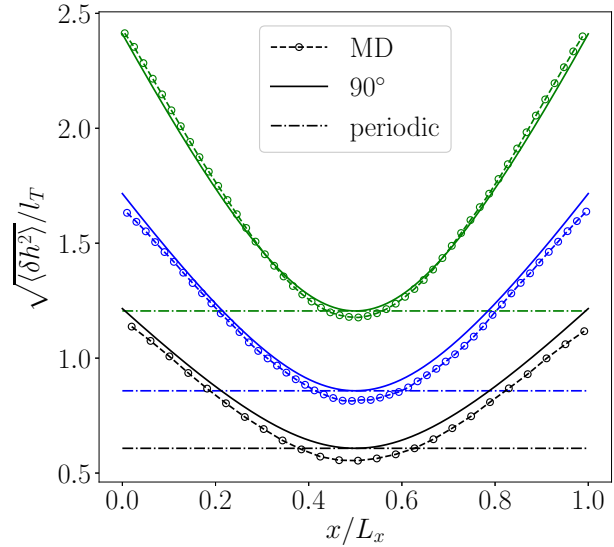}% Here is how to import EPS art
\caption{\label{fig:meniscus_dh2} Standard deviation of the fluctuations of films with $90^\circ$ contact angle (black: $L_x=13.04$ nm, blue: $L_x=25.99$, green: $L_x=51.29$ nm). MD results (dashed lines with circles) are compared to our theory, Eq.~(\ref{eq:TCW_90}) (solid lines). Results are normalized by the thermal length scale $l_T$. The dashed-and-dotted horizontal lines are fluctuation amplitudes predicted by Eq.~(\ref{eq:TCW_periodic}) for a periodic (unbounded) film.} 
\end{figure}

Fig.~\ref{fig:meniscus_dh2} shows the standard deviation of the free surface fluctuations, normalized by the thermal length scale $l_T$, obtained from MD simulations and compared to our theoretical predictions, Eq.~(\ref{eq:TCW_90}); the agreement is excellent. The fluctuation amplitudes of an unbounded film (i.e. adopting a periodic boundary condition, Eq.~(\ref{eq:TCW_periodic})) are also provided. The relative strength of thermal fluctuations of the film interface increase with film length, as expected \cite{vinkCapillaryWavesColloidpolymer2005,delgado-buscalioniHydrodynamicsNanoscopicCapillary2008}. However, comparing to Eq.~(\ref{eq:TCW_periodic}) we can see that our expression predicts an enhanced fluctuation amplitude to that of a periodic film everywhere except at the center, $x=L_x/2$, where they coincide.  Physically, this is because the replacement of periodicity with a fixed contact angle permits additional (`half') wave modes, i.e. of the form $\cos((2n-1)\pi x/L_x)$ for $n=1,2,...$, which contribute to a larger amplitude everywhere except at $x=L_x/2$, where they are zero. Another interesting observation is that the effects of boundaries propagate across the whole film, regardless of the film length.

\subsection{Partially pinned contact lines}
\label{sec:2d_pinned}
Now we turn our attention to the case where the contact lines are pinned onto the walls. The position of contact lines can be restrained by chemical heterogeneity \cite{wenMolecularDropletsVs2021} or physical defects \cite{nadkarniInvestigationMicroscopicAspects1992,schafferDynamicsContactLine1998}. However, in MD it is not possible to perfectly pin the interface at a height $h=h_0$, as thermal fluctuations cause it to fluctuate, even if just mildly around the target pinning height. Therefore, to compare MD and theory, we must account for this and do so by modeling the contact line as a Langevin diffusion process; as done in \cite{fernandez-toledanoContactlineFluctuationsDynamic2019}. Then, the `partially' pinned boundary condition can be written as
\begin{equation}
\label{eq:BCLangevin}
    h(0,t) = N_1(t)+h_0,\qquad  h(L_x,t) = N_2(t)+h_0.
\end{equation}
Here $N_1(t)$ and $N_2(t)$ are Langevin diffusion processes governed by
\begin{align}
\label{eq:Langevin}
	\xi\frac{dN_1}{dt} &= -kN_1(t) + f_1(t),\\
	\xi\frac{dN_2}{dt} &= -kN_2(t) + f_2(t),
\end{align}
where $\xi$ is the so-called coefficient of friction, $k$ is the harmonic constant, and $f_1(t)$ and $f_2(t)$ are Gaussian noise functions that satisfy $\langle f_1(s)f_1(\tau)\rangle=2\xi k_B T \delta(s-\tau)$ and $\langle f_2(s)f_2(\tau)\rangle=2\xi k_B T \delta(s-\tau)$. 

From this model, the correlation of $N$ has the form \cite{kampenStochasticProcessesPhysics2007}
\begin{equation}
\label{eq:Ncorr}
    \langle N(s)N(\tau)\rangle = \frac{k_B T}{k}e^{-\frac{k}{\xi}|s-\tau|},
\end{equation}
and when $s=\tau$ Eq.~(\ref{eq:Ncorr}) simply gives the variance of $N$ as
\begin{equation}
\label{eq:varLangevin}
    \langle N^2\rangle = \frac{k_B T}{k}.
\end{equation}
By fitting the exponential curve of Eq.~(\ref{eq:Ncorr}) and the variance Eq.~(\ref{eq:varLangevin}) to MD simulations data we can calculate $k$ and $\xi$. 

Our problem in this case is then solving the STFE Eq.~(\ref{eq:2dSLE}) with the partially-pinned-contact-line condition Eq.~(\ref{eq:BCLangevin}) and the impermeable side-wall condition Eq.~(\ref{eq:BCnoflux}).

\subsubsection{Bulk modes}

\begin{figure}[t]
    \includegraphics[width=\linewidth]{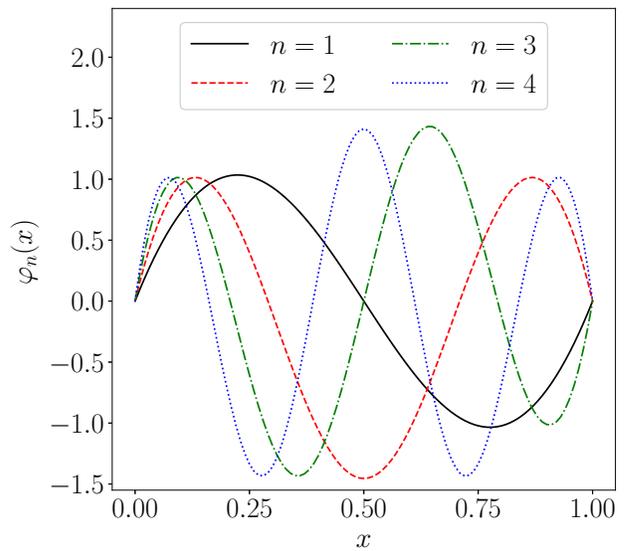}
    \caption{Wave modes $\varphi_n(x)$ for a film with perfectly pinned contact lines. }
    \label{fig:varphi}
\end{figure}

For a \textit{perfectly} pinned contact line, the appropriate wave modes (see Appendix \ref{app:modes_pinned}) are
\begin{eqnarray}
	\varphi_n(x) &&= \sinh(\lambda_n^{1/4}x)+\sin(\lambda_n^{1/4}x)\nonumber\\
	&&+ K\big(\cosh(\lambda_n^{1/4}x)-\cos(\lambda_n^{1/4}x)\big)
\end{eqnarray}
with eigenvalues
\begin{equation}
    \lambda_n\approx \left(\frac{\pi/2 + n\pi}{L_x}\right)^4,\qquad n=1,2,\ldots
\end{equation}
As distinct from the $90^\circ$-contact-angle case, the mode corresponding to the $\lambda_0=0$ case also exists:
\begin{equation}
	\varphi_0(x) = x\left(1-\frac{x}{L_x}\right).
\end{equation}
Although $\varphi_n(x)$ are not orthogonal, for odd $n$ they are odd functions around $x=L_x/2$ and for even $n$ they are even functions around $x=L_x/2$. We will exploit this property to simplify the calculation for fluctuation amplitudes later on. Figure~\ref{fig:varphi} provides an illustration of $\varphi_n(x)$.

\subsubsection{Decomposition of fluctuations}

The partially-pinned-contact-line boundary condition is a linear combination of the perfectly pinned condition and the Langevin diffusion condition. This suggests that under linearization the free surface can be decomposed as
\begin{equation}
\label{eq:pinned_dcomp}
    h(x,t) = h_0 + h_2(x,t) + h_3(x,t),
\end{equation}
where $h_0$ is the initial position of the contact line and $h_2(x,t)$, $h_3(x,t)$ are small perturbations. Applying this to Eq.~(\ref{eq:2dSLE}), at the leading order ($h_2\sim h_3 \sim \mathcal{N}\ll h_0$) we obtain
\begin{equation}
    \frac{\partial h_2}{\partial t} + \frac{\partial h_3}{\partial t} = -\frac{\gamma h_0^3}{3\mu}\left(\frac{\partial^4 h_2}{\partial x^4}+\frac{\partial^4 h_3}{\partial x^4}\right)+\sqrt{\frac{2k_B Th_0^3}{3\mu L_y}}\frac{\partial \mathcal{N}}{\partial x}
\end{equation}
and the boundary conditions in Eq.~(\ref{eq:BCnoflux}) and Eq.~(\ref{eq:BCLangevin}) become
\begin{align}
    h_2(0,t)&+h_3(0,t) = N_1(t),\\
    h_2(L_x,t) &+ h_3(L_x,t) = N_2(t),\\
    \frac{\partial^3 h_2}{\partial x^3}(0,t) &+ \frac{\partial^3 h_3}{\partial x^3}(0,t) = 0,\\
    \frac{\partial^3 h_2}{\partial x^3}(L_x,t) &+ \frac{\partial^3 h_3}{\partial x^3}(L_x,t) = 0.
\end{align}
This is actually a linear combination of two smaller problems, one with a noise-driven bulk and pinned contact lines
\begin{subequations}
\label{eq:noiseBulk}
\begin{align}
    &\frac{\partial h_2}{\partial t} = -\frac{\gamma h_0^3}{3\mu}\frac{\partial^4 h_2}{\partial x^4} + \sqrt{\frac{2k_B Th_0^3}{3\mu L_y}}\frac{\partial \mathcal{N}}{\partial x},\\
    &h_2(0,t) = h_2(L_x,t) = 0,\\
    &\frac{\partial^3 h_2}{\partial x^3}(0,t) = \frac{\partial^3 h_2}{\partial x^3}(L_x,t) = 0,
\end{align}
\end{subequations}
and the other with deterministic equations in the bulk and noise-driven contact lines
\begin{subequations}
\label{eq:noiseBoundary}
\begin{align}
    &\frac{\partial h_3}{\partial t} = -\frac{\gamma h_0^3}{3\mu}\frac{\partial^4 h_3}{\partial x^4},\\
    &h_3(0,t) = N_1(t),\: h_3(L_x,t) = N_2(t),\\
    &\frac{\partial^3 h_3}{\partial x^3}(0,t) = \frac{\partial^3 h_3}{\partial x^3}(L_x,t) = 0.
\end{align}
\end{subequations}

We can then solve Eqs.~(\ref{eq:noiseBulk}) for $h_2(x,t)$ by decomposing it into wave modes $\varphi_n(x)$ 
\begin{equation}
\label{eq:pinned_h2}
    h_2 = \sum_{n=1}^{\infty} c_n(t)\varphi_n(x),
\end{equation}
where $c_n(t)$ are wave amplitudes that can be expressed explicitly (see  Appendix \ref{app:pinned_dh2_SLE}). Similarly $h_3(x,t)$ can also be decomposed into wave modes $\varphi_n(x)$ and boundary modes,
\begin{align}
\notag
    h_3(x,t) &= \sum_{n=1}^N e_n(t)\varphi_n(x)\\
    \notag
    &+ N_1(t)\left(1-\frac{x}{L_x}\right)\left(1-\frac{x}{L_x}-u_{10}\frac{x}{L_x}\right)\\
    \label{eq:2d_pinned_h3}
    &+ N_2(t)\frac{x}{L_x}\left[\frac{x}{L_x}-u_{20}\left(1-\frac{x}{L_x}\right)\right].
\end{align}
where $e_n(t)$ are wave amplitudes with explicit expressions, and $u_{10}$ and $u_{20}$ are constants that are given in Appendix \ref{app:SolveLangevin}. Note, only $N$ wave modes are considered for $h_3(x,t)$, opposed to infinitely many wave modes considered for $h_2(x,t)$. This is because the wave modes are not orthogonal to each other, so when solving the linear system for $e_n(t)$, matrix $G$ is non-diagonal and it would be impossible to take its inverse if the dimension is infinite (see Appendix \ref{app:SolveLangevin}). We can confirm numerically that this does not affect our results (the fluctuation amplitudes converge) and in Section~\ref{sec:cutoff} we show that a cut-off on the number of wave modes is actually preferable.

\subsubsection{Thermal-capillary-wave theory}

We can obtain the fluctuation amplitude for $h_2$ using thermal-capillary-wave theory. Similar to the $90^\circ$-contact-angle case, we substitute Eq.~(\ref{eq:pinned_dcomp}) into Eq.~(\ref{eq:TCW}) and use the fact that the $\partial\varphi_n/\partial x$ are orthogonal (see Appendix \ref{app:modes_pinned}) to obtain
\begin{equation}
\label{eq:2d_pinned_energy}
    E = \frac{\gamma L_x L_y}{2}\sum_{n=1}^{\infty}\lambda_n^{1/2}c_n^2 + \hbox{other terms},
\end{equation}
where the first term on the right hand is the change of surface area due to $h_2$ and the other terms are the change of surface area due to $h_3$ and cross terms of $h_2$ and $h_3$. Applying the equipartition theorem to the $h_2$ only terms we find
\begin{equation}
\label{eq:equipartitionPinned}
    \langle c_n^2\rangle = l_T^2\frac{1}{L_xL_y}\frac{1}{\lambda_n^{1/2}}.
\end{equation}
Using the fact that $\langle c_m c_n\rangle=\delta_{mn}\langle c_n^2\rangle$ (see Appendix \ref{app:pinned_dh2_SLE}), finally, we have
\begin{equation}
\label{eq:2d_pinned_dh22}
    \langle h_2^2(x)\rangle = l_T^2\frac{1}{L_xL_y}\sum_{n=1}^{\infty}\frac{\varphi_n^2(x)}{\lambda_n^{1/2}}.
\end{equation}

Note, alternatively one can derive $\langle c_n^2\rangle$ directly from the STFE, which includes time dependence,
\begin{equation}
    \langle c_n^2\rangle = \frac{k_B T}{\gamma}\frac{1}{L_x L_y}\frac{1}{\lambda^{1/2}}(1-\exp(-2C\lambda_n t)),
\end{equation}
where $C=\gamma h_0^3/(3\mu)$ (see Appendix~\ref{app:pinned_dh2_SLE}). At thermal equilibrium this agrees with Eq.~(\ref{eq:equipartitionPinned}).

\subsubsection{Combined fluctuation amplitude}

The fluctuations combine to give a total variance of 
\begin{equation}
\label{eq:TCW_pin}
    \langle (h_2+h_3)^2\rangle = \langle h_2^2\rangle + 2\langle h_2h_3\rangle + \langle h_3^2\rangle.
\end{equation}
where $\langle h_2^2\rangle$ is the fluctuation of the bulk, already calculated via thermal-capillary-wave theory, and $\langle h_3^3 \rangle$ is the fluctuation of the film originating from fluctuations of the contact lines. Notably, since the random variables $\mathcal{N}(x,t)$, $f_1(t)$ and $f_2(t)$ are uncorrelated, $\langle h_2h_3\rangle=0$ (see Appendix \ref{app:Langevin_dh2_SLE}).
An expression for $\langle h_3^2(x)\rangle$, obtained from Eq.~(\ref{eq:apph3FA}), can be found in Appendix \ref{app:Langevin_dh2_SLE}.

\subsubsection{Molecular-dynamics simulations}
\label{sec:2d_pinned_MD}
\begin{figure}[t]
    \includegraphics[width=\linewidth]{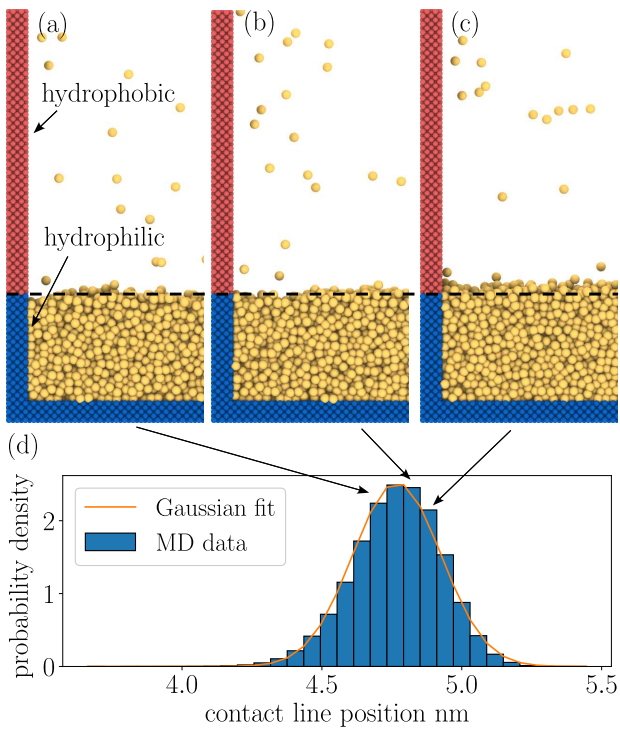}
    \caption{(a), (b) and (c) are MD snapshots in a region near the contact line and the chemical heterogeneity of the solid wall, showing the position of the contact line below, on and above the proposed pinning point respectively. Blue particles denote hydrophilic wall atoms and red particles denote the hydrophobic wall atoms. Liquid Argon atoms are in yellow. (d) shows a histogram of contact line position extracted from a single realization of MD simulations. }
    \label{fig:MD_2d_pin}
\end{figure}

\begin{table}[b]
\caption{\label{tab:pinned}Simulation parameters and their non-dimensional values (reduced units based on Lennard-Jones potential parameters) for pinned contact lines.
}
\begin{ruledtabular}
\begin{tabular}{cccc}
Property & Nondim.value & Value & Unit  \\
\hline\\[-2ex]
$\epsilon_{sf1}$ & 0.05  & $0.0835\times 10^{-21}$ & J  \\
$\epsilon_{sf2}$ & 0.62 & $1.0354\times 10^{-21}$ & J  \\
$\sigma_{sf1}$ & 0.8 & $0.272$ & nm  \\
$\sigma_{sf2}$ & 0.8 & $0.272$ & nm  \\
\end{tabular}
\end{ruledtabular}
\end{table}

The MD simulations are the same as in Section~\ref{sec:2d_90_MD} with the exception that we need to pin the contact line. There are several ways to achieve this, for example, by using topographical defects on the solid substrate \cite{gleasonMicrodropletEvaporationForced2014}, but here we use the technique described by Kusudo \textit{et. al}  \cite{kusudoExtractionEquilibriumPinning2019} using chemical heterogeneity. As shown in Fig.~\ref{fig:MD_2d_pin}, this is achieved by using a hydrophilic wall (blue) beneath the film's equilibrium height ($h_0$) and a hydrophobic one (red), which is less wettable, above it. The wettability of the walls are tuned by changing the interaction parameters between solid and liquid, $\epsilon_{sf1}$ and $\epsilon_{sf2}$ \cite{fernandez-toledanoContactlineFluctuationsDynamic2019}. This results in the position of the contact line following a Gaussian distribution with mean $h_0$, when in thermal equilibrium, as can be seen from Fig.~\ref{fig:MD_2d_pin} (d). The variance depends on the equilibrium contact angles (i.e. on the $\epsilon_{sf}$'s) of the walls and a small variance is preferable to mimic perfect pinning. Our choice of parameters are shown in Table~\ref{tab:pinned}.

Four different film lengths are tested: Film 4 ($L_x=13.04$ nm), Film 5 ($L_x=25.99$ nm), Film 6 ($L_x=51.29$ nm) and Film 7 ($L_x=102.30$ nm). The film width $L_y=2.94$ nm and the initial film height $h_0=4.85$ nm are the same as in the $90^\circ$-contact-angle case. The equilibration time $t_c$ can be estimated from Eq.~(\ref{eq:apppinEquT}). Multiple independent MD simulations are performed (Film 4: $18$, Film 5: $10$, Film 6: $10$ and Film 7: $20$).

\begin{figure}[h!]
\includegraphics[width=\linewidth]{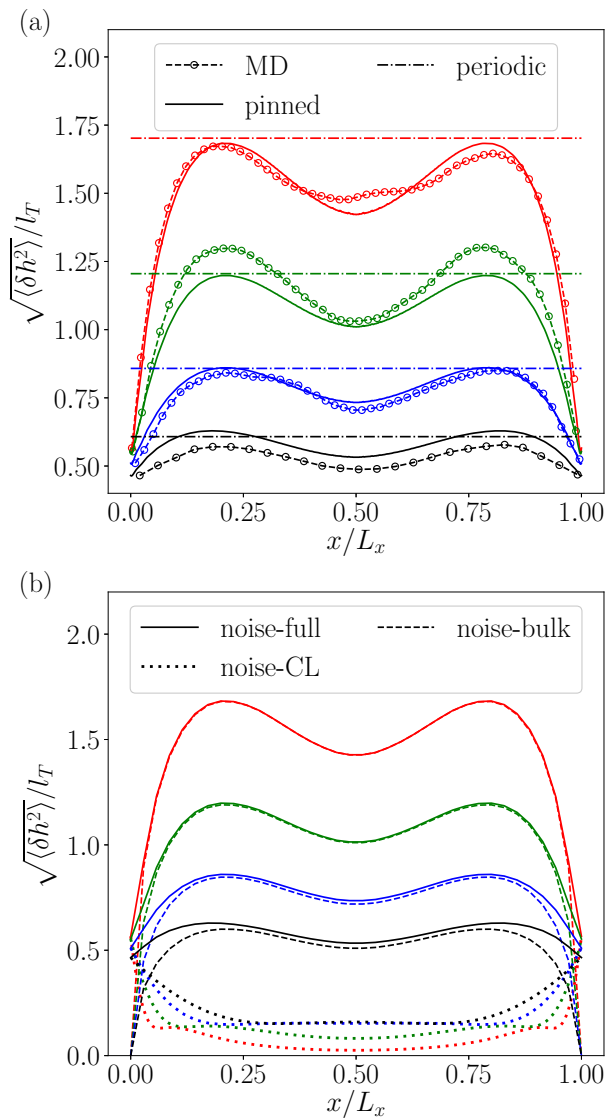}
\caption{\label{fig:pinned_dh2} Standard deviation of fluctuations for the films with partially pinned contact lines (black: $L_x=13.04$ nm, blue: $L_x=25.99$ nm, green: $L_x=51.29$ nm, red: $L_x=102.30$ nm): (a) shows a comparison of MD (dashed lines with circles), theory via Eq.~(\ref{eq:TCW_pin}) (solid lines) and the classic thermal-capillary-wave theory for periodic films (dashed lines) predicted by Eq.~(\ref{eq:TCW_periodic}); (b) shows the decomposition of theory Eq.~(\ref{eq:TCW_pin}), where the dashed lines represent the amplitudes of fluctuations caused by thermal noise in bulk, the dotted lines represent the amplitudes of fluctuations caused by noise on the contact lines and the solid lines represent the full fluctuation amplitudes.}
\end{figure}

Fig.~\ref{fig:pinned_dh2}(a) shows the fluctuation amplitudes of the free surface obtained from MD simulations using bins with side length $1.5\sigma_{f\!f}$ in the $x$-direction and $1.4\sigma_{f\!f}$ in the $y$-direction, which agree well with the theoretical predictions, Eq.~(\ref{eq:TCW_pin}). Notably, the largest difference between the MD and theory occurs for the shortest films; an effect we will revisit in Section~\ref{sec:cutoff}. One can see that the fluctuation amplitudes of films with partially pinned contact lines have a saddle shape with one trough and two crests, symmetric about $x=L_x/2$ as we would expect. In contrast to the previous case of a fixed contact angle, here the fluctuations are almost everywhere lower than those for a periodic film. The amplitudes dip significantly at the wall, due to the pinning effect, but do not reach zero due to the oscillations around the pinning position. Moreover, the positions of the trough and the crests relative to the length of the film are fixed, showing again that the boundary effects propagate across the film. Lastly, as suggested by our theory Eq.~(\ref{eq:TCW_pin}), the fluctuation amplitudes of the free surface can be attributed to the thermal noise in the bulk $\sqrt{\langle h_2^2\rangle}$ and the fluctuations of the contact line $\sqrt{\langle h_3^3\rangle}$. From the decomposition of fluctuation amplitudes shown in Fig.~\ref{fig:pinned_dh2} (b) one can see that the effect of contact line fluctuations is limited to the region near the boundaries and clearly in this regime the bulk fluctuations are stronger than the contact-line-driven motions.

\section{3D circular Bounded thin films}
\label{sec:3d}

\begin{figure}
    \centering
    \includegraphics[width=\linewidth]{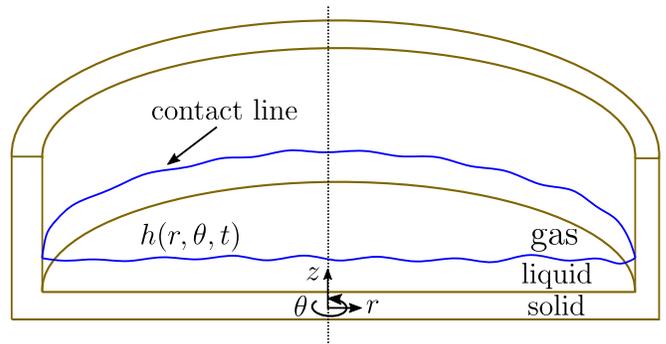}
    \caption{An illustration of the geometry of the 3D circular thin films (half to show cross section).}
    \label{fig:3dthinfilm}
\end{figure}

Let us now extend our investigation to three-dimensional bounded nanofilms. In 3D, the position of the free surface $h(\mathbf{x},t)$ is given by the thin-film equation (TFE) \cite{morrisonIntroductionFluidMechanics2013} as
\begin{equation}
    \label{eq:3dLE}
    \frac{\partial h}{\partial t} = -\frac{\gamma}{3\mu}\nabla\cdot\left(h^3\nabla\nabla^2h\right),
\end{equation}
and its stochastic version \cite{davidovitchSpreadingViscousFluid2005,grunThinFilmFlowInfluenced2006,zhangNanoscaleThinfilmFlows2020} is
\begin{equation}
    \label{eq:3dSLE}
    \frac{\partial h}{\partial t} = -\frac{\gamma}{3\mu}\nabla\cdot\left(h^3\nabla\nabla^2h\right)+\sqrt{\frac{2k_B T}{3\mu}}\nabla\cdot\left(h^{3/2}\boldsymbol{\mathcal{N}}\right),
\end{equation}
with thermal noise uncorrelated in space and time:
\begin{equation}
\label{eq:3dnoise_property}
    \langle \mathcal{N}_i(\mathbf{x},t)\mathcal{N}_j(\mathbf{x}',t')\rangle = \delta_{ij} \delta(\mathbf{x}-\mathbf{x}')\delta(t-t').
\end{equation}

As in quasi-2D, thermal fluctuations drive nanowaves on the free surface. When periodic boundary conditions (i.e. an unconfined film) are considered on a square domain of length $L$ the perturbation $\delta h$ to the average film height can be decomposed into Fourier modes and the fluctuation amplitude is given by
\begin{align}
\label{eq:3d_dh2_per_nocf}
    \langle \delta h^2\rangle &=\frac{l_T^2}{\pi^2}\sum_{m=1}^{\infty}\sum_{n=1}^{\infty}\frac{1}{m^2+n^2},
\end{align}
where $m$ and $n$ are wavenumbers in the $x$-direction and the $y$-direction. Unlike quasi-2D, this summation is unbounded and therefore an upper limit to the wavenumbers $m$ and $n$ is required. A natural choice is to consider a `cut-off' length scale $\ell_c$, such that wave modes with length scale less than $\ell_c$ are ignored 
\cite{flekkoyFluctuatingFluidInterfaces1995,flekkoyFluctuatingHydrodynamicInterfaces1996,grantFluctuatingHydrodynamicsCapillary1983}
\begin{align}
\label{eq:3d_dh2_per}
    \langle \delta h^2\rangle &=\frac{l_T^2}{\pi^2}\sum_{m=1}^{m<L/\ell_c}\sum_{n=1}^{n<L/\ell_c}\frac{1}{m^2+n^2},
\end{align}
A discussion on the significance of the requirement for a cut-off length scale is provided in Section~\ref{sec:cutoff}. If $L/\ell_c\gg 1$, which is not unreasonable given $\ell_c$ will be on the molecular scale, the summation Eq.~(\ref{eq:3d_dh2_per}) can be approximated by
\begin{equation}
     \langle \delta h^2\rangle \approx l_T^2 \ln\frac{L}{\ell_c},
\end{equation}
showing a logarithmic growth of the fluctuation amplitude with the length of the domain $L$. This growth, although much slower than the linear growth in Eq.~(\ref{eq:TCW_periodic}) for the quasi-2D periodic boundary case, is nevertheless unbounded as $L\to\infty$.  

Let us now consider how the analysis is modified when we have confined 3D films. In particular, we choose to confine liquid films in circular domains by solid walls as illustrated in Fig.~\ref{fig:3dthinfilm}. As in quasi-2D, we apply two different boundary conditions: (i) $90^\circ$ contact angle and (ii) pinned contact line. 

\subsection{Prescribed Angle at $90^\circ$}

It is natural to conduct the analysis in cylindrical coordinates $(r,\theta,z)$, with a thin film $h=h(r,\theta,t)$ of equilibrium height $h_0$ confined by an impermeable wall at $r=a$. Then, a prescribed $90^\circ$ contact angle corresponds to
\begin{equation}
    \label{eq:3dBC90}
    \left.\frac{\partial h}{\partial r}\right|_{r=a} = 0, \qquad \theta\in[0,2\pi).
\end{equation}
and impermeability of the wall is
\begin{equation}
    \label{eq:3dBCnoflux}
    \left.\nabla\nabla^2 h\right|_{r=a}\cdot \hat{\bm{r}} = 0,\qquad \theta\in[0,2\pi).
\end{equation}
The impermeability condition corresponds to a projection of the flux onto the direction normal to the wall, given by $\hat{\bm{r}}$, which is a unit vector in $r$.

\begin{figure}[t]
\includegraphics[width=\linewidth]{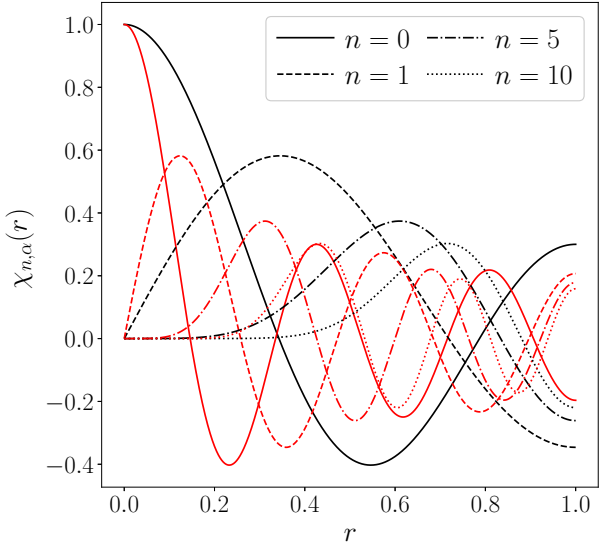}% Here is how to import EPS art
\caption{\label{fig:chi} Wave modes $\chi_{n,\alpha}(r)$ for the circular-bounded 3D film with $90^\circ$ contact angle. $n$ is the wave number in $\theta$ and $\alpha$ is the wave number in $r$. $\alpha=2$ for black lines and $\alpha=5$ for red lines.}
\end{figure}

Linearizing the 3D TFE, Eq.~(\ref{eq:3dLE}), and solving the eigenvalue problem with the above boundary conditions (Eq.~(\ref{eq:3dBC90}) and Eq.~(\ref{eq:3dBCnoflux})), we obtain the following wave modes: (see Appendix \ref{app:3dmodes_90})
\begin{align}
    \Upsilon^1_{n,\alpha}(r,\theta) &= \cos(n\theta)\chi_{n,\alpha}(r),\\
    \Upsilon^2_{n,\alpha}(r,\theta) &= \sin(n\theta)\chi_{n,\alpha}(r).
\end{align}
Here
\begin{equation}
    \chi_{n,\alpha}(r) = J_n(\omega_{n,\alpha} r),\qquad n=0,1,\ldots,
\end{equation}
are the wave modes in $r$. $J_n$ is the $n$th Bessel function of the first kind. The dispersion relation
\begin{equation}
\label{eq:3d90dispersion}
    J'_n(\omega a) = 0, \qquad n=0,1,\ldots
\end{equation}
where $'$ denotes a derivative, is derived from solving the eigenvalue problem; from the dispersion relation we obtain the frequencies $\{\omega_{n,\alpha}:\alpha=1,2,\ldots\}$. Fig.~\ref{fig:chi} gives an illustration of $\chi_{n,\alpha}(r)$ with different $n$ and $\alpha$. One can see that the $90^\circ$-contact-angle condition is satisfied. As $n$ increases the position of the first crest gets further away from the origin and there is an expanded region in which the wave mode's amplitude is negligible.  

\subsubsection{Thermal-capillary-wave theory}
\label{sec:3d90TCW}
The free surface can be written in terms of the wave modes:
\begin{equation}
\label{eq:3D90decomp}
    h(r,\theta,t) = h_0 + h_4(r,\theta,t),
\end{equation}
where the perturbation $h_4$ is
\begin{align}
\notag
    h_4(r,\theta,t) = \sum_{n=0}^\infty\sum_{\alpha=1}^\infty &\Big[A_{n,\alpha}(t)\Upsilon^1_{n,\alpha}(r,\theta)\\
    &+B_{n,\alpha}(t)\Upsilon^2_{n,\alpha}(r,\theta) \Big].
\end{align}
Then we can calculate the energy of a perturbed surface (see Appendix \ref{app:3dTCW})
\begin{align}
\notag
    E &= \gamma\Bigg(\int_0^{2\pi}\int_0^a\sqrt{1+\left(\frac{\partial h}{\partial r}\right)^2+\frac{1}{r^2}\left(\frac{\partial h}{\partial \theta}\right)^2}rdrd\theta-\pi a^2\Bigg)\\
    \notag
    &\approx \frac{\gamma}{2}\int_0^{2\pi}\int_0^a\left(\left(\frac{\partial h}{\partial r}\right)^2+\frac{1}{r^2}\left(\frac{\partial h}{\partial \theta}\right)^2\right)rdrd\theta\\
    \label{eq:3d_90_energy}
    &=\gamma\pi\sum_{\alpha=1}^{\infty}A_{0,\alpha}^2 S_{0,\alpha}+\frac{\gamma\pi }{2}\sum_{n=1}^{\infty}\sum_{\alpha=1}^{\infty}(A_{n,\alpha}^2+B_{n,\alpha}^2)S_{n,\alpha},
\end{align}
where 
\begin{equation}
    S_{n,\alpha} = \frac{1}{2}(\omega_{n,\alpha}^2 a^2-n^2)J_n^2(\omega_{n,\alpha}a).
\end{equation}
Since the wave modes are orthogonal to each other (see Appendix \ref{app:3dTCW}) and the energy is quadratic in the amplitudes, we can use the equipartition theorem to give
\begin{equation}
\label{eq:3d_90_equi1}
    \frac{k_B T}{2} = \gamma S_{0,\alpha}\langle A_{0,\alpha}^2\rangle
\end{equation}
and
\begin{equation}
\label{eq:3d_90_equi2}
    \frac{k_B T}{2} = \frac{\gamma \pi}{2}S_{n,\alpha}\langle A_{n,\alpha}^2\rangle = \frac{\gamma \pi}{2}S_{n,\alpha}\langle B_{n,\alpha}^2\rangle,
\end{equation}
so that the (position-dependent) variance of fluctuations is given by
\begin{align}
\notag
    \langle h_4^2(r,\theta) \rangle &= \frac{k_B T}{\gamma}\frac{1}{\pi}\Big[\sum_{\alpha=1}^{\infty}\frac{1}{2S_{0,\alpha}}\chi_{0,\alpha}^2(r)\\
    \label{eq:3d_dh2_90_nocut}
    &+\sum_{n=1}^{\infty}\sum_{\alpha=1}^{\infty}\frac{1}{S_{n,\alpha}}\chi_{n,\alpha}^2(r)\Big].
\end{align}
Note that the variance is only $r$ dependent, since periodicity in $\theta$ eliminates variations.

Asymptotic analysis shows that $S_{n,\alpha}$ increases with $n$ linearly (see Appendix~\ref{app:3dTCW}). So the summation over $n$ in Eq.~(\ref{eq:3d_dh2_90_nocut}) diverges as $n\to\infty$ and a cut-off for smallest length scale $\ell_c$ should be introduced, as we've already seen for the periodic (i.e. unbounded) 3D film Eq.~(\ref{eq:3d_dh2_per}). 

Applying a cut-off length scale in polar coordinates is non-trivial, as the modes in the $r$ and the $\theta$ directions differ, in contrast to the Cartesian case, and the radial wavemodes have non-trivial form. To do so, we define the length scale of a wave mode $f_{n,\alpha}(r,\theta)$ as
\begin{equation}
\label{eq:3d_90_LS}
    \mathcal{L}(f_{n,\alpha}) = \frac{\max_{r\in[0,a],\theta\in[0,2\pi]}|f_{n,\alpha}(r,\theta)|}{\max_{r\in[0,a],\theta\in[0,2\pi]}\left|\nabla f_{n,\alpha}(r,\theta)\right|},
\end{equation}
where $||$ is the absolute value and $\nabla$ is the gradient operator. For simplicity, we denote the length scale of the wave mode for the $90^\circ$-contact-angle case as $L^{90}_{n,\alpha} = \mathcal{L}(\Upsilon^1_{n,\alpha})$, and one can easily check that $\mathcal{L}(\Upsilon^1_{n,\alpha})=\mathcal{L}(\Upsilon^2_{n,\alpha})$. We then introduce a threshold function 
\begin{equation}
    Z^{90}(\ell_c,n,\alpha)=
    \begin{cases}
        1,&\text{ if }L_{n,\alpha}^{90}\geq\ell_c\\
        0,&\text{ if }L_{n,\alpha}^{90}<\ell_c
    \end{cases},
\end{equation}
to identify wave modes with length scale greater than a chosen $\ell_c$. Finally, we apply the cut-off to Eq.~(\ref{eq:3d_dh2_90_nocut})
\begin{align}
\notag
    \langle h_4^2(r,\theta) \rangle &= \frac{k_B T}{\gamma}\frac{1}{\pi}\Big[\sum_{\alpha=1}^{\infty}\frac{Z^{90}(\ell_c,0,\alpha)}{2S_{0,\alpha}}\chi_{0,\alpha}^2(r)\\
    \label{eq:3d_dh2_90_cut}
    &+\sum_{n=1}^{\infty}\sum_{\alpha=1}^{\infty}\frac{Z^{90}(\ell_c,n,\alpha)}{S_{n,\alpha}}\chi_{n,\alpha}^2(r)\Big]
\end{align}
which regularizes the unbounded sum. The effect of cut-offs will be further discussed in Section~\ref{sec:cutoff}.

\subsubsection{Molecular-dynamics simulations}
\label{sec:3d_90_MD}

\begin{figure}[t]
\includegraphics[width=\linewidth]{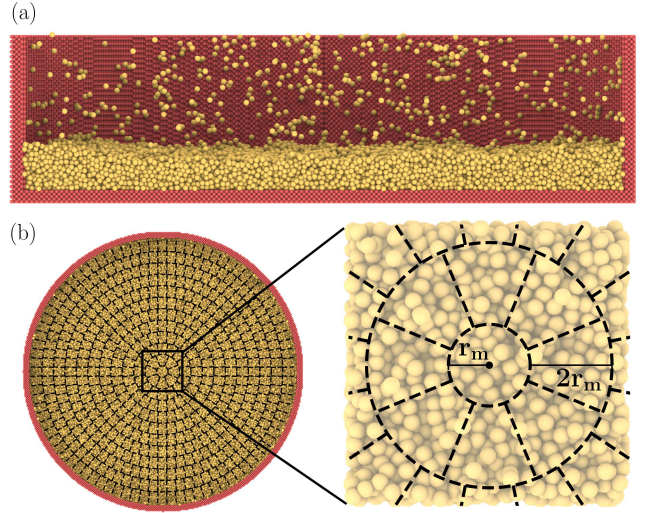}% Here is how to import EPS art
\caption{\label{fig:circular_mesh} Snapshots of MD simulations for a 3D circular film: (a) shows the cross section, (b) shows the top view with the circular mesh used for extracting the free surface position.}
\end{figure}

The setup of the MD simulations is very similar to before, except for geometry. The cylindrical side wall is joined by a circular base, both $5$ layers of Platinum atoms in fcc, to form a `cup'. An equilibrated $h_0=2.5$ nm thick Argon liquid film is then placed on top of the base with a $0.17$ nm gap, as in Section~\ref{sec:2d_90_MD}. Equilibrated Argon vapor is then placed on top of the liquid. Non-periodic boundary conditions with reflective walls are applied at the top, as a result the vapor can not escape the cup. Fig.~\ref{fig:circular_mesh} (a) gives a snapshot of the MD simulation.

The position of the free surface is measured using the number density and binning technique, with a circular mesh used to reduce errors near the wall. Calculation of number density and identification of liquid Argon particles are the same as for quasi-2D. To define the vertical bins, we: (i) choose the number of layers $N_l$, (ii) create a center bin as a circle with radius $r_m=a/(2N_l-1)$ and area $\pi r_m^2$, (iii) ensure bins in the other layers are rings with width of $2r_m$, (iv) divide each ring equally into tiles such that the area of each tile is also $\pi r_m^2$. Fig.~\ref{fig:circular_mesh} (b) shows an illustration of the circular mesh with $N_L=12$. The characteristic length scale of the mesh is given by the square root of the area of the tiles  $L_m=\sqrt{\pi r_m^2}$.

The parameters for the MD simulations are still given in Table~\ref{tab:90} and we confirm that the average contact angle remains at $90$ degrees. Films with two different radii are tested: Film 8 ($a=11.81$ nm) and Film 9 ($a=23.39$ nm). The average height $h_0$ of the free surface is measured to be $2.47$ nm.

\begin{figure}[h]
\includegraphics[width=\linewidth]{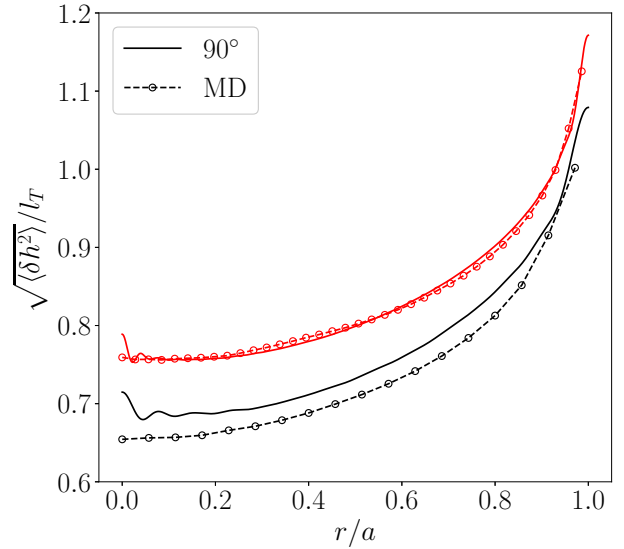}% Here is how to import EPS art
\caption{\label{fig:3d_meniscus_dh2} Standard deviation of fluctuations for $90^\circ$-contact-angle 3D circular films with two different radii (black: $a=11.81$ nm, red: $a=23.39$ nm). MD simulation results (dashed lines with circles) and theory Eq.~(\ref{eq:3d_dh2_90_cut}) (solid lines) are normalized by the thermal scale $l_T$. }
\end{figure}

The fluctuation amplitudes extracted from MD simulations are averaged over $\theta$ since they are only $r$ dependent. Fig.~\ref{fig:3d_meniscus_dh2} shows the fluctuation amplitudes of 3D circular bounded films with a $90^\circ$ contact angle (normalized by the thermal length scale $l_T$) obtained from MD simulations and compared with the theoretical prediction Eq.~(\ref{eq:3d_dh2_90_cut}). The smallest length scale allowed in the theory is chosen to be $\ell_c=\sigma_{f\!f}$ and the length scale of the circular mesh for MD is chosen proportionally $L_m=1.77\ell_c$ (i.e. $r_m=\sigma_{f\!f}$). One can see that the MD results agree well with Eq.~(\ref{eq:3d_dh2_90_cut}) for both films, while the agreement improves as the film gets larger. The MD results indicate that similar to the quasi-2D films with $90^\circ$ contact angle, the minimum of the fluctuation amplitude is found at the center of the film. This is because the first crest of wave modes for $n\geq 1$ get pushed further away from the origin as $n$ increases, shown in Fig.~(\ref{fig:chi}), distributing less energy to the center and more energy towards the boundary. One can also observe oscillations in the theory near the origin, which is absent in MD simulation results. This could indicate that a better cut-off mechanism is needed near the singularity ($r=0$), or MD resolutions may need to be increased to capture the oscillation; both worthy of future investigation.

\subsection{Pinned contact lines}
\begin{figure}[t]
\includegraphics[width=\linewidth]{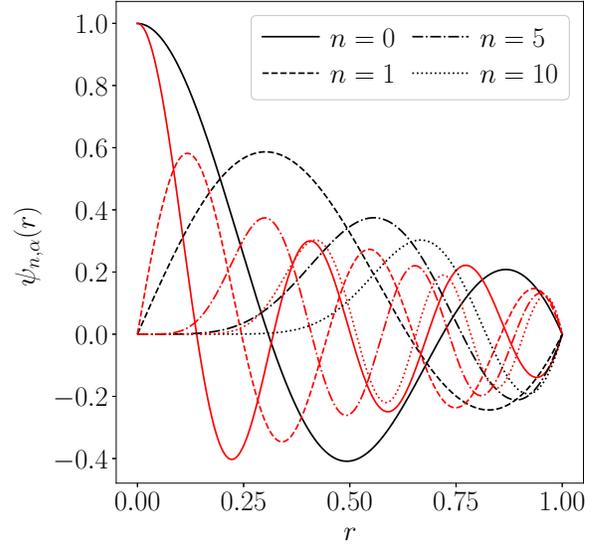}% Here is how to import EPS art
\caption{\label{fig:psi} Wave modes $\psi_{n,\alpha}(r)$ for the circular bounded 3D film with a pinned contact line. $n$ is the wave number in $\theta$ and $\alpha$ is the wave number in $r$. $\alpha=2$ for black lines and $\alpha=5$ for red lines.}
\end{figure}

Next, we consider the case where the contact line is pinned onto the wall. As mentioned in Section~\ref{sec:2d_pinned}, in practice the contact line will always oscillate in MD simulations. However, due to the complexity in 3D, and the relatively less prominent influence of contact line fluctuations previously observed, the theory we develop will only consider the contact line being pinned perfectly onto the wall. It will be interesting in future work to  explore the oscillation of the contact line in 3D. The pinned-contact-line boundary condition for the 3D circular film is given by
\begin{equation}
\label{eq:3dBCpinned}
    h(a,\theta) = h_0,\qquad \theta\in[0,2\pi].
\end{equation}
Together with Eq.~(\ref{eq:3dLE}) and Eq.~(\ref{eq:3dBCnoflux}) we can obtain the appropriate wave modes (see Appendix \ref{app:3dmodes_pinned})
\begin{align}
\label{eq:3d_pinned_modes}
    &\Psi^1_{n,\alpha}(r,\theta) = \cos(n\theta)\psi_{n,\alpha}(r),\\
    &\Psi^2_{n,\alpha}(r,\theta) = \sin(n\theta)\psi_{n,\alpha}(r).
\end{align}
Here
\begin{align}
\notag
    \psi_{n,\alpha}(r) = J_n(\zeta_{n,\alpha} r)-\frac{J_n(\zeta_{n,\alpha} a)}{I_n(\zeta_{n,\alpha} a)}&I_n(\zeta_{n,\alpha} r),\\
    & n=0,1,\ldots
\end{align}
are the wave modes in $r$. $I_n$ is the $n$th modified Bessel function of first kind. The frequencies $\{\zeta_{n,\alpha}:\alpha=1,2,\ldots\}$ are obtained from a dispersion relation
\begin{align}
    \notag
    2n J_n(\zeta a)I_n(\zeta a) &+ \zeta a\Big[J_n(\zeta a)I_{n+1}(\zeta a)\\
    &-J_{n+1}(\zeta a)I_n(\zeta a)\Big] = 0,
\end{align}
derived from the eigenfunction problem. Fig.~\ref{fig:psi} gives an illustration of $\psi_{n,\alpha}(r)$ with different $n$ and $\alpha$. The pinned boundary condition is satisfied and the distance between the origin and the first crest still increases with $n$.

\subsubsection{Thermal-capillary-wave theory}
If we perturb the free surface from the mean profile $h_0$ we obtain
\begin{equation}
    h(r,\theta,t) = h_0 + h_5(r,\theta,t)
\end{equation}
where the perturbation $h_5(r,\theta,t)$ can be decomposed into wave modes:
\begin{align}
\notag
    h_5(r,\theta,t) = \sum_{n=0}^\infty\sum_{\alpha=1}^\infty &\Big[C_{n,\alpha}(t)\Psi^1_{n,\alpha}(r,\theta)\\
    &+D_{n,\alpha}(t)\Psi^2_{n,\alpha}(r,\theta) \Big].
\end{align}
Following the same procedure as before, we find that the energy required to perturb the surface is given by (details see Appendix \ref{app:3dpinnedTCW})
\begin{equation}
\label{eq:3d_pinned_energy}
    E = \gamma\pi\sum_{\alpha=1}^{\infty}A_{0,\alpha}^2 K_{0,\alpha}+\frac{\gamma\pi }{2}\sum_{n=1}^{\infty}\sum_{\alpha=1}^{\infty}(C_{n,\alpha}^2+D_{n,\alpha}^2)K_{n,\alpha},
\end{equation}
where
\begin{align}
\notag
    K_{n,\alpha} &=\frac{1}{2}\omega_{n,\alpha}^2a^2\Bigg(-J_{n-1}(\omega_{n,\alpha}a)J_{n+1}(\omega_{n,\alpha}a)\\
    &+I_{n-1}(\omega_{n,\alpha}a)I_{n+1}(\omega_{n,\alpha}a)\frac{J_n^2(\omega_{n,\alpha}a)}{I_n^2(\omega_{n,\alpha}a)}\Bigg)
\end{align}
Assuming the wave modes are uncorrelated, we can apply the equipartition theorem:
\begin{equation}
    \frac{k_B T}{2} = \gamma K_{0,\alpha}\langle C_{0,\alpha}^2\rangle,
\end{equation}
\begin{equation}
    \frac{k_B T}{2} = \frac{\gamma \pi}{2}K_{n,\alpha}\langle C_{n,\alpha}^2\rangle = \frac{\gamma \pi}{2}K_{n,\alpha}\langle D_{n,\alpha}^2\rangle,
\end{equation}
to find the variance of the fluctuations:
\begin{align}
    \langle h_5^2(r,\theta) \rangle &= \frac{k_B T}{\gamma}\frac{1}{\pi}\Big[\sum_{\alpha=1}^{\infty}\frac{1}{2K_{0,\alpha}}\psi_{0,\alpha}^2(r)\\
    &+\sum_{n=1}^{\infty}\sum_{\alpha=1}^{\infty}\frac{1}{K_{n,\alpha}}\psi_{n,\alpha}^2(r)\Big].
\end{align}
The value of $K_{n,\alpha}$ increases with $n$ linearly (see Appendix~\ref{app:3dpinnedTCW}) so that, as expected, the fluctuation amplitude diverges and a cut-off for length scale $\ell_c$ should be introduced. The length scales of the wave modes for the pinned case are again calculated by Eq.~(\ref{eq:3d_90_LS}) and now denoted by $L^p_{n,\alpha} = \mathcal{L}(\Psi^1_{n,\alpha}) = \mathcal{L}(\Psi^2_{n,\alpha})$. Introducing the threshold function
\begin{equation}
    Z^{p}(\ell_c,n,\alpha)=
    \begin{cases}
        1,&\text{ if }L_{n,\alpha}^{p}\geq\ell_c\\
        0,&\text{ if }L_{n,\alpha}^{p}<\ell_c
    \end{cases},
\end{equation}
and we can write the regularized sum as
\begin{align}
\notag
    \langle h_5^2(r,\theta) \rangle &= \frac{k_B T}{\gamma}\frac{1}{\pi}\Big[\sum_{\alpha=1}^{\infty}\frac{Z^p(\ell_c,0,\alpha)}{2K_{0,\alpha}}\psi_{0,\alpha}^2(r)\\
    \label{eq:3d_dh2_pin_cut}
    &+\sum_{n=1}^{\infty}\sum_{\alpha=1}^{\infty}\frac{Z^p(\ell_c,n,\alpha)}{K_{n,\alpha}}\psi_{n,\alpha}^2(r)\Big].
\end{align}
The choice of cut-offs will be further discussed in Section~\ref{sec:cutoff}.

\subsubsection{Molecular-dynamics simulations}

The geometry of the MD simulations is set and the position of the free surface is measured using the same methods described in Section~\ref{sec:3d_90_MD}. MD parameters from Table~\ref{tab:pinned} are used. The rest of the MD settings are the same as in Section~\ref{sec:2d_90_MD}. Films with two different radii are tested: Film 10 ($a=11.81$ nm) and Film 11 ($a=23.39$ nm).

\begin{figure}[h]
\includegraphics[width=\linewidth]{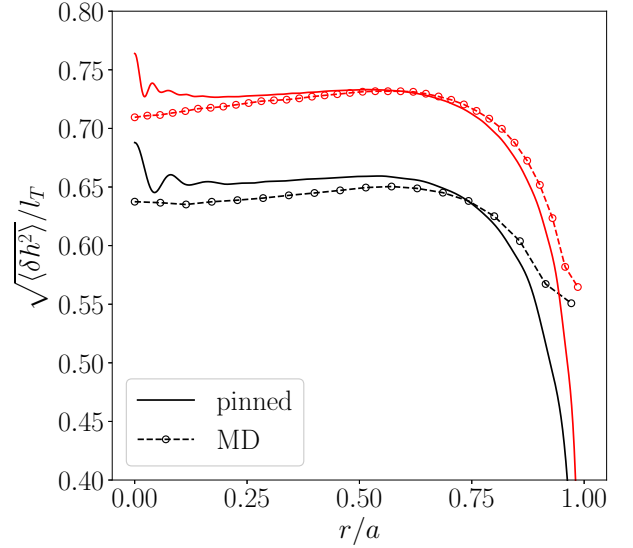}% Here is how to import EPS art
\caption{\label{fig:3d_pinned_dh2} Standard deviation of fluctuations for pinned-contact-line 3D circular films with two different radii (black: $a=11.81$ nm, red: $a=23.39$ nm). MD simulation results (dashed lines with circles) and theory Eq.~(\ref{eq:3d_dh2_pin_cut}) (solid lines) are normalized by thermal scale $l_T$. }
\end{figure}

Fig.~\ref{fig:3d_pinned_dh2} shows the spatial variation of fluctuation amplitudes of the 3D circular films with a pinned contact line. The figure compares the theoretical predictions of the fluctuation amplitudes Eq.~(\ref{eq:3d_dh2_pin_cut}) (by setting cut off length scale $\ell_c=\sigma_{f\!f}$) with the MD results (obtained with a circular binning mesh of characteristic length scale $L_m=1.77\ell_c$) ; again the overall agreement is very good, with improvements for larger films. The theoretical predictions also exhibit oscillations near the origin, whereas the MD results do not. This might suggest a singularity in the solution is requiring better cut-off mechanism, or a low MD resolution failed to detect the oscillations, as stated in Section~\ref{sec:3d_90_MD}. Similar to the quasi-2D films with partially pinned contact lines, the MD results exhibit a trough at the center ($r=0$ and $x=L_x/2$) and a crest before reaching the boundary.

\section{Discussion on minimum length scales}
\label{sec:cutoff}

We have seen that our theoretical models for 3D films require a minimum length scale to be defined in order for predictions to be made; a length-scale ‘cut-off’ is needed.
For the results presented, where we have compared to Molecular Dynamics, this cut-off was chosen on a physical basis, coinciding with the Lennard-Jones length-scale parameter $\sigma_{f\!f}$. Willis \textit{et. al} \cite{willisThermalCapillaryWaves2010} observed, in MD simulation, rapid attenuation of the fluctuation strength of thermal capillary waves (on a 2D film) at scales beneath $\sigma_{f\!f}$, and so this was a natural first choice.

\begin{figure}[h!]
\includegraphics[width=\linewidth]{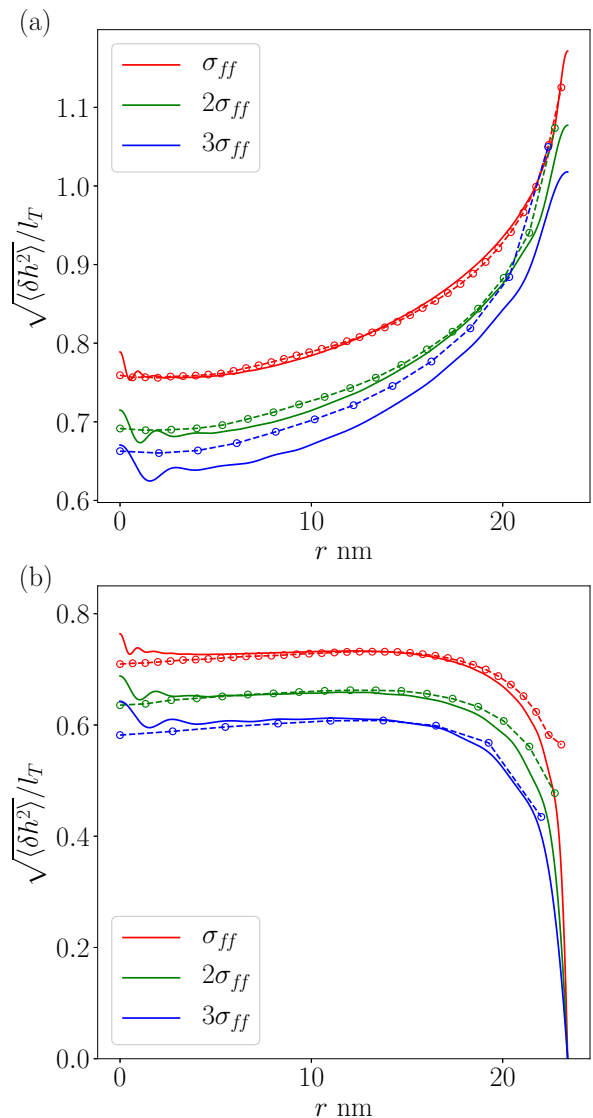}
\caption{\label{fig:3d_dh2_cutoff} Standard deviation of fluctuation of 3D circular film with radius $a=23.39$ nm and different boundary conditions: (a) $90^\circ$ contact angle and (b) a pinned contact line. Theoretical predictions Eq.~(\ref{eq:3d_dh2_90_cut}) Eq.~(\ref{eq:3d_dh2_pin_cut}) with different cut-off length scales $\ell_c=\sigma_{f\!f}$, $2\sigma_{f\!f}$, $3\sigma_{f\!f}$ are given by solid lines. The circled lines are MD results obtained using different bin sizes $L_m=1.77\ell_c$. }
\end{figure}

However, in the case of MD, there is another length scale that might influence the results: the bin size over which measurements are spatially averaged. A more sophisticated choice of cut-off for our theoretical model should, then, be either bin size or $\sigma_{f\!f}$, whichever is larger. Up to now, coincidentally, they have been equal. To test if predictions from MD are indeed modified by the bin size when it is larger than $\sigma_{f\!f}$, and that our theoretical model with an appropriately adjusted cut-off predicts this, we have performed the data processing presented in Figure~\ref{fig:3d_dh2_cutoff}. It shows that, indeed, the overall fluctuation strength of the film in MD is influenced by the choice of bin size, and that this is well captured by the theory with a bin-size cut-off. Unfortunately, we were unable to perform simulations with bin sizes smaller than $\sigma_{f\!f}$, as done in Willis \textit{et. al} \cite{willisThermalCapillaryWaves2010}, where we might expect the effect of bin size to disappear, revealing a minimum scale comparable with $\sigma_{f\!f}$. We leave this for clarification in follow-on work.

\begin{figure}[h!]
\includegraphics[width=\linewidth]{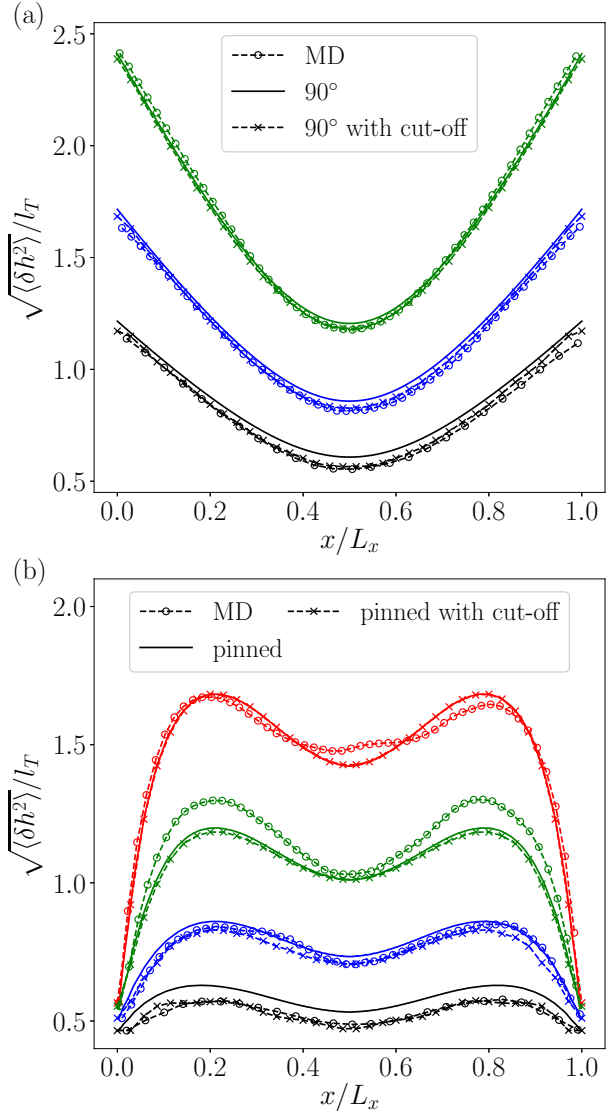}
% Here is how to import EPS art
\caption{\label{fig:2d_dh2_cutoff} Comparison of fluctuation amplitudes for quasi-2D thin films extracted from MD (circled lines), theoretical predictions considering a cut-off (crossed lines) and without a cut-off (solid lines). (a) $90^\circ$ contact angle, (b) partially pinned contact lines. Black: $L_x=13.04$ nm, blue: $L_x=25.99$ nm, green: $L_x=51.29$ nm, red: $L_x=102.30$ nm. }
\end{figure}

As these results indicate, be it physical ($\sigma_{f\!f}$) or numerical (bin size), the results from MD are affected by a minimum length scale. While we originally introduced the ‘cut-off’ out of necessity for a bounded sum in our theoretical model for 3D films, this discussion implies that introducing a cut-off in the theoretical model for quasi-2D films should still improve the comparison with MD. This comparison is made in Figure~\ref{fig:2d_dh2_cutoff}, and indeed there is a small but noticeable improvement in agreement.

\section{Conclusion \& Future Directions}
\label{app:future}
In this article, we have uncovered the behavior of confined nanoscale films in thermal equilibrium. These results, in particular the spatial dependence of the fluctuation amplitude, could be validated experimentally, using either scattering techniques \cite{liCoupledCapillaryWave2001,huObservationLowviscosityInterface2006,alvineCapillaryWaveDynamics2012} or colloid-polymer mixtures to enable optical measurement \cite{aartsDirectVisualObservation2004} -- quasi-2D results could be approximated using Hele-Shaw type geometries whilst 3D domains are the norm. Furthermore, the techniques used could be extended to tackle a range of other nanoscale flows including free films, drops or bubbles. 

Our findings serve to further highlight the accuracy of fluctuating hydrodynamics to describe nanoscale fluid phenomena, or, put another way, to reproduce effects seen in molecular simulations at a fraction of the computational cost. Moreover, the results presented could provide useful benchmarks for computational schemes intended at describing nanoscale flows and give insight into the choice of cut-off used to regularize the singular noise terms in the stochastic partial differential equations; which can be achieved either using projections onto regular bases \cite{duran-olivenciaInstabilityRuptureFluctuations2019,grunThinFilmFlowInfluenced2006} or just be crudely based on the numerical grid size.

Notably, in many cases one is interested primarily in the stability of nanovolumes. For thin films \cite{beckerComplexDewettingScenarios2003}, the importance of thermal fluctuations has been established \cite{fetzerThermalNoiseInfluences2007} but the relation to nano-confinement is yet to be determined; this could also be a direction of future research.

\begin{acknowledgments}
This work was supported by the EPSRC grants EP/W031426/1, EP/S029966/1, EP/S022848/1,  EP/P031684/1, EP/V01207X/1, EP/N016602/1 and the NSFC grant 12202437. Jingbang Liu is supported by a studentship within the UK Engineering and Physical Sciences Research Council–supported Centre for Doctoral Training in modeling of Heterogeneous Systems, Grant No. EP/S022848/1. The authors are grateful to Dr Ed Brambley for discussions on the orthogonality of eigenfunctions. 
\end{acknowledgments}

\appendix

\section{Quasi-2D thin film with $90^\circ$ contact angle}
In this section we layout the technical details for the quasi-2D thin film with $90^\circ$ contact angle.
\subsection{Derivation of the wave modes}
\label{app:modes_90}
The surface wave can no longer be decomposed into Fourier modes since the boundary conditions are not periodic. To find appropriate wave modes, we first linearize the thin-film equation Eq.~(\ref{eq:2dLE}) and then solve the eigenvalue problem. Consider a perturbation to the free surface of the form
\begin{equation}
	h(x,t) = h_0 + \epsilon h_1(x)T(t),
\end{equation}
where we anticipate that the perturbation is separable in time $T(t)$ and space $h_1(x)$, the steady state is a flat free surface $h=h_0$ and $\epsilon\ll 1$. Apply this to the thin-film equation (\ref{eq:2dLE}), at the leading order we obtain a linear problem
\begin{equation}
\label{eq:app90linODE}
	\frac{1}{T}\frac{dT}{dt} = -\frac{\gamma h_0^3}{3\mu h_1}\frac{d^4 h_1}{dx^4} = -\omega,
\end{equation}
where $\omega$ is a constant and must be positive for stability. This gives an eigenvalue problem for $h_1(x)$ with corresponding boundary conditions
\begin{equation}
	\frac{d^4 h_1}{dx^4} = \sigma h_1,\qquad \sigma = \frac{3\mu\omega}{\gamma h^3_0}\geq 0
\end{equation}
\begin{equation}
    \left.\frac{d h_1}{dx}\right|_{x=0}=\left.\frac{d h_1}{dx}\right|_{x=L_x}=\left.\frac{d^3 h_1}{dx^3}\right|_{x=0}=\left.\frac{d^3 h_1}{dx^3}\right|_{x=L_x}.
\end{equation}
Solving the eigenvalue problem gives the appropriate wave modes
\begin{equation}
    \phi_n(x) = \cos\left((\sigma_n)^{1/4} x\right),\: n=1,2,\ldots,
\end{equation}
where the associated eigenvalues are
\begin{equation}
    \sigma_n = \left(\frac{n\pi}{L_x}\right)^{4},\: n=1,2,\ldots
\end{equation}
Solving Eq.~(\ref{eq:app90linODE}) for $T(t)$ we get
\begin{equation}
    T = T_0 \exp(-\omega t) = T_0 \exp\left(-\sigma\frac{\gamma h_0^3}{3\mu}t\right),
\end{equation}
which gives us an estimate of how fast the perturbation decay and how long it takes for the wave modes to equilibrate. The wave mode with longest length scale takes longest to equilibrate
\begin{equation}
\label{eq:app90equT}
    t_e \approx \frac{3\mu}{\gamma h_0^3 \sigma_1}.
\end{equation}

In this subsection we (i) derived the wave modes for quasi-2D $90^\circ$-contact-angle case and (ii) evaluated the time for the wave modes to equilibriate, which guides us to the runtime of MD simulations.

\subsection{Fluctuation amplitude from the STFE}
\label{app:90_dh2_SLE}
Applying Eq.~(\ref{eq:90_dcomp}) to the STFE Eq.~(\ref{eq:2dSLE}), at the leading order we obtain
\begin{equation}
	\sum_{n=1}^{\infty}\phi_n\frac{d a_n}{dt} = -\frac{\gamma h_0^3 \pi^4}{3\mu L_x^4}\sum_{n=1}^{\infty}n^4a_n\phi_n + \sqrt{\frac{2k_BTh_0^3}{3\mu L_y}}\frac{\partial \mathcal{N}}{\partial x}.
\end{equation}
The noise is then expanded in the wave modes $\bar{\phi}_n = \sin(n\pi x/L_x)$, so that
\begin{equation}
	\mathcal{N}(x,t) = \sum_{m=1}^{\infty} b_m(t)\bar{\phi}_m(x)
\end{equation}
and using the orthogonality of the $\bar{\phi}$'s and noting that $\int_0^L \bar{\phi}_m^2dx=L_x/2$ we find
\begin{equation}
	\label{eq:b_n_property}
	b_m = \frac{2}{L_x}\int_0^{L_x}\bar{\phi}_m \mathcal{N}dx.
\end{equation}
This allows us to write an equation for each mode
\begin{equation}
\label{eq:app2d90factor}
	\phi_n\frac{d a_n}{dt} = -C n^4a_n\phi_n + Db_n\frac{d\bar{\phi}_n}{dx},
\end{equation}
where we have introduced constants $A = \frac{\gamma h_0^3 \pi^4}{3\mu L_x^4}$ and $B=\sqrt{\frac{2k_BTh_0^3}{3\mu L_y}}$. We can then rewrite Eq.~(\ref{eq:app2d90factor}) using an integrating factor to find
\begin{equation}
	\phi_n\frac{d}{dt}\left(a_ne^{An^4t}\right) = Be^{An^4 t}b_n\frac{d\bar{\phi}_n}{dx}.
\end{equation}
Integrating both sides with time and assuming that the initial film is flat, i.e. $a_n(0)=0$, we have
\begin{equation}
	\phi_n a_n = Be^{-An^4t}\frac{d\bar{\phi}_n}{dx}\int_0^te^{An^4\tau}b_n(\tau)d\tau.
\end{equation}
and noting $\frac{d\bar{\phi}_n}{dx}=\frac{n\pi}{L_x}\phi_n$, we find
\begin{equation}
	a_n(t) = \frac{Bn\pi}{L_x}e^{-An^4t}\int_0^te^{An^4\tau}b_n(\tau)d\tau.
\end{equation}

Next, using Eq.~(\ref{eq:2dnoise_property}) and Eq.~(\ref{eq:b_n_property}) we determine the properties of the noise coefficients
\begin{align}
\notag
	\langle b_n(\tau)b_n(s)\rangle &= \langle \frac{4}{L_x^2}\left(\int_0^{L_x}\bar{\phi}_n(x)\mathcal{N}(x,\tau)dx\right)\\
	\notag
	&\times\left(\int_0^{L_x}\bar{\phi}_n(x')\mathcal{N}(x',s)dx'\right)\rangle \\
	\notag
	&=\frac{4}{L_x^2}\int_0^{L_x}\int_0^{L_x}\bar{\phi}_n(x)\bar{\phi}_n(x')\\
	\notag
	&\times\delta(x-x')\delta(\tau-s)dxdx'\\
	&=\frac{2}{L_x}\delta(\tau-s), 
\end{align}
from which we finally obtain
\begin{align}
\notag
	\langle a_n^2 \rangle &= \frac{2B^2n^2\pi^2e^{-2An^4t}}{L_x^3}\int_0^te^{2An^4\tau}d\tau\\
	&=\frac{B^2\pi^2(1-e^{-2An^4t})}{An^2L_x^3}.
\end{align}
This gives a time dependent version of $\langle a_n^2\rangle$, and as $t\to\infty$ (i.e. we approach thermal equilibrium) we have
\begin{equation}
	\langle a_n^2 \rangle = \frac{2k_B T}{\gamma\pi^2}\frac{L_x}{L_y}\frac{1}{n^2},
\end{equation}
which agrees with the result from thermal-capillary-wave theory Eq.~(\ref{eq:equipartition90}).

We can also show that $\langle a_m a_n\rangle=\delta_{mn}\langle a_n^2\rangle$,
\begin{eqnarray}
\langle a_m(t) &&a_n(t)\rangle = De^{-C(m^4+n^4)t}\nonumber\\
&&\times \int_0^t\int_0^te^{Cm^4s+Cn^4\tau}\langle b_m(s)b_n(\tau)\rangle dsd\tau
\end{eqnarray}
where
\begin{widetext}
\begin{align}
    \notag
	\langle b_m(s)b_n(\tau)\rangle &= \langle \frac{4}{L_x^2}\left(\int_0^{L_x}\bar{\phi}_m(x)\mathcal{N}(x,s)dx\right)\left(\int_0^{L_x}\bar{\phi}_n(x')\mathcal{N}(x',\tau)dx'\right)\rangle \\
	\notag
	&=\langle\frac{4}{L_x^4}\int_0^{L_x}\int_0^{L_x}\mathcal{N}(x,s)\mathcal{N}(x',\tau)\bar{\phi}_m(x)\bar{\phi}_n(x')dxdx'\rangle\\
	\notag
	&=\frac{4}{L_x^2}\int_0^{L_x}\int_0^{L_x}\langle \mathcal{N}(x,s)\mathcal{N}(x',\tau)\rangle\bar{\phi}_m(x)\bar{\phi}_n(x')dxdx'\\
	\notag
	&=\frac{2\delta(s-\tau)}{L_x^2}\int_0^L\bar{\phi}_m(x')\bar{\phi}_n(x')dx'\\
	&=\frac{2\delta(s-\tau)}{L_x}\delta_{mn}.
\end{align}
\end{widetext}
This shows that the wave modes are uncorrelated and is required in the calculation of Eq.~(\ref{eq:TCW_90}).

In this subsection we showed that (i) the amplitudes of the wave mode $\langle a_n^2\rangle$ can be derived directly from the STFE as a function of time, which agrees with thermal-capillary-wave theory at thermal equilibrium (as $t\to\infty$) and (ii) the wave mode are uncorrelated, which is an requirement for the calculation in Eq.~(\ref{eq:TCW_90}).

\section{Quasi-2D thin film with partially pinned contact lines}
In this section we layout the technical details for quasi-2D thin film with partially pinned contact lines.
\subsection{Derivation of wave modes}
\label{app:modes_pinned}

The partially pinned boundary condition given by Eq.~(\ref{eq:BCLangevin}) states that the contact-lines oscillates around the pinned point $h_0$. However, first we derive the wave modes with perfectly pinned boundary condition with the contact-lines pinned at $h_0$.  After the same linearization as in Appendix \ref{app:modes_90} we arrive at the eigenvalue problem with pinned and no-flux boundary conditions
\begin{equation}
	\frac{d^4 h_2}{dx^4} = \lambda h_2,
\end{equation}
\begin{equation}
    h_2(0)=h_2(L_x)=\frac{d^3 h_2}{dx^3}(0)=\frac{d^3 h_2}{dx^3}(L_x)=0.
\end{equation}
The eigenvalue problem gives a general solution \cite{courantMethodsMathematicalPhysics2009}
\begin{eqnarray}
	h_2(x) &&= C_1\cosh(\lambda^{1/4}x) + C_2\sinh(\lambda^{1/4}x)\nonumber\\
	&&+C_3\cos(\lambda^{1/4}x)+C_4\sin(\lambda^{1/4}x),
\end{eqnarray}
where $C_1$-$C_4$ are constants to be determined. Substituting in the boundary conditions gives the appropriate wave modes
\begin{eqnarray}
	\varphi_n(x) &&= \sinh(\lambda_n^{1/4}x)+\sin(\lambda_n^{1/4}x)\nonumber\\
	&&+ K\big(\cosh(\lambda_n^{1/4}x)-\cos(\lambda_n^{1/4}x)\big)
\end{eqnarray}
where
\begin{equation}
    K = \frac{\bigg(\sinh(\lambda_n^{1/4}L_x)+\sin(\lambda_n^{1/4}L_x)\bigg)}{\bigg(\cos(\lambda_n^{1/4}L_x)-\cosh(\lambda_n^{1/4}L_x)\bigg)}.
\end{equation}
The eigenvalues $\lambda_n$ must satisfy
\begin{equation}
	\cosh(\lambda^{1/4}_nL_x)\cos(\lambda^{1/4}_nL_x) = 1,
\end{equation}
which gives us an estimate
\begin{equation}
    \lambda_n\approx \left(\frac{\pi/2 + n\pi}{L_x}\right)^4,\qquad n=1,2,\ldots
\end{equation}
And for $\lambda_0=0$, we have
\begin{equation}
	\varphi_0(x) = \frac{x}{L_x}\left(1-\frac{x}{L_x}\right).
\end{equation}
Similar to Appendix (\ref{app:modes_90}) we can estimate the equilibration time $t_c$ by looking at the wave mode with longest length scale apart from $0$
\begin{equation}
\label{eq:apppinEquT}
    t_c= \frac{3\mu}{\gamma h_0^3 \lambda_1}.
\end{equation}

Although the wave modes are not orthogonal, their first derivatives are, so that we are able to make analytic progress. To show this, let $'$ denote $\partial/\partial x$, then using integration by parts repeatedly we can show
\begin{align}
	\lambda_m\int_0^{L_x}&\varphi_m'\varphi_n'dx = \int_0^{L_x}\varphi_m'''''\varphi_n'dx\nonumber\\ 
	&= [\varphi_m''''\varphi_n']_0^{L_x}+\int_0^{L_x}\varphi_m''''\varphi_n''dx \nonumber \\
	&=[\varphi_m''''\varphi_n'-\varphi_m'''\varphi_n'']_0^{L_x}+\int_0^L\varphi_m'''\varphi_n'''dx\nonumber\\
	&=[\varphi_m''''\varphi_n'-\varphi_m'''\varphi_n''+\varphi_m''\varphi_n''']_0^{L_x}\nonumber\\
	&-\int_0^{L_x}\varphi_m''\varphi_n''''dx\nonumber\\
	&=[\varphi_m''''\varphi_n'-\varphi_m'''\varphi_n''+\varphi_m''\varphi_n'''-\varphi_m'\varphi_n'''']_0^{L_x}\nonumber\\
	&+\int_0^{L_x}\varphi_m'\varphi_n'''''dx\nonumber\\
	&=\lambda_n\int_0^{L_x}\varphi_m'\varphi_n'dx,
\end{align}
where all the boundary terms vanish due to boundary conditions. Since $\lambda_m\neq\lambda_n$, we have shown that $\varphi'$ are orthogonal.

In this subsection we (i) derived the wave modes $\varphi_n$ for quasi-2D thin films with partially pinned contact lines, (ii) derived the time for the wave modes to reach equilibrium, and (iii) showed that $\varphi'_n$ are orthogonal, which is used in derivation of Eq.~(\ref{eq:2d_pinned_energy}).

\subsection{Fluctuation amplitude from the STFE}
\label{app:pinned_dh2_SLE}
Similar to before, we would like to derive the mean square displacement directly from the stochastic thin-film equation for $h_2$. Considering a perturbation and expanding it in the derived wave modes gives
\begin{equation}
\label{eq:appDh2}
	h = h_0 + \sum_{n=0}^{\infty} c_n(t)\varphi_n(x).
\end{equation}
Using similar arguments and noting that $\varphi_n^{(4)} = \lambda_n\varphi_n$ we find
\begin{equation}
\label{eq:pinned_linear_SLE}
	\sum_{n=0}^{\infty}\varphi_n\frac{dc_n}{dt} = -\frac{\gamma h_0^3}{3\mu}\sum_{n=1}^{\infty}\lambda_n c_n\varphi_n + \sqrt{\frac{2k_BTh_0^3}{3\mu L_y}}\frac{\partial \mathcal{N}}{\partial x},
\end{equation}
Note that $\lambda_0=0$, this is why the second term begins with $n=1$.
To continue we need to expand the noise in some basis as well. Since there is already a first derivative in space on $\mathcal{N}$ we expand the noise with $\bar{\varphi}_n=\varphi_n'''$. Note that $\varphi_0'''=0$, so $\bar{\varphi}_0=0$. So we can write the noise in terms of the basis
\begin{equation}
	\mathcal{N}(x,t) = \sum_{m=1}^{\infty} d_m(t)\bar{\varphi}_m(x)
\end{equation}
and using the orthogonality of $\bar{\varphi}_m$ and noting that $\int_0^{L_x}\bar{\varphi}_m^2dx=L_x$ we have
\begin{equation}
	\label{eq:b_n_property_pinned}
	d_m = \frac{1}{L_x}\int_0^{L_x}\bar{\varphi}_m\mathcal{N}dx.
\end{equation}
We can then rewrite equation (\ref{eq:pinned_linear_SLE}) in each mode as
\begin{equation}
	\varphi_n\frac{dc_n}{dt} = -C\lambda_nc_n\varphi_n + Dd_n\frac{d\bar{\varphi}_n}{dx},
\end{equation}
where we have introduced new constants $C=\frac{\gamma h_0^3}{3\mu}$ and $D = \sqrt{\frac{2k_B Th_0^3}{3\mu L_y}}$. For $\varphi_0$ we have
\begin{equation}
	\frac{dc_0}{dt}=0,
\end{equation} 
and since the average free surface is flat we have $c_0 = 0$. We can then solve the ordinary differential equations for $n\neq 0$ as before to get
\begin{equation}
	\varphi_n c_n = De^{-C\lambda_n t}\frac{d\bar{\varphi}_n}{dx}\int_0^te^{C\lambda_n\tau}d_n(\tau)d\tau.
\end{equation}
Noting $\frac{d\bar{\varphi}_n}{dx}=\lambda^{1/4}\varphi_n$, we have
\begin{equation}
\label{eq:appDcn}
	c_n = D\lambda_n^{1/4}e^{-C\lambda_nt}\int_0^te^{C\lambda_n\tau}d_n(\tau)d\tau.
\end{equation}
Following equation (\ref{eq:2dnoise_property}) and (\ref{eq:b_n_property_pinned}) we find that
\begin{align}
\notag
	\langle d_n(\tau)d_n(s)\rangle &= \langle \frac{1}{L_x^2}\left(\int_0^{L_x}\bar{\varphi}_n(x)\mathcal{N}(x,\tau)dx\right)\\
	\notag
	&\times\left(\int_0^L\bar{\varphi}_n(x')\mathcal{N}(x',s)dx'\right)\rangle \\
	\notag
	&=\frac{1}{L_x^2}\int_0^{L_x}\int_0^{L_x}\bar{\varphi}_n(x)\bar{\varphi}_n(x')\\
	\notag
	&\times\delta(x-x')\delta(\tau-s)dxdx'\\
	&=\frac{1}{L_x}\delta(\tau-s).
\end{align}
So we can write
\begin{align}
\notag
	\langle c_n^2\rangle &= \langle D^2\lambda_n^{1/2}e^{-2C\lambda_n t}\int_0^te^{C\lambda_n\tau}d_n(\tau)d\tau\\
	\notag
	&\times\int_0^te^{C\lambda_n s}d_n(s)ds\rangle \\
	\notag
	&= \frac{D^2(1-e^{-2C\lambda_nt})}{2L_xC\lambda_n^{1/2}} \\
	&= \frac{k_BT}{\gamma}\frac{1}{L_x L_y}\frac{1}{\lambda_n^{1/2}}(1-e^{-2C\lambda_nt}),
\end{align}
and as $t\to\infty$
\begin{equation}
	\langle c_n^2\rangle = \frac{k_BT}{\gamma}\frac{1}{L_x L_y}\frac{1}{\lambda_n^{1/2}}
\end{equation}

Similarly we can also show that
\begin{equation}
	\langle d_m(s)d_n(\tau)\rangle = \frac{\delta(s-\tau)}{L_x}\delta_{mn},
\end{equation}
and so
\begin{align}
\notag
	\langle c_m(t) c_n(t)\rangle &= \langle D^2\lambda_m^{1/4}\lambda_n^{1/4}e^{-C(\lambda_m+\lambda_n)t}\\
	\notag
	&\times\int_0^t\int_0^t e^{C\lambda_ms+C\lambda_n\tau}d_m(s)d_n(\tau)dsd\tau\rangle \\
	\notag
	&= \frac{D^2\lambda_m^{1/4}\lambda_n^{1/4}}{L_x}e^{-C(\lambda_m+\lambda_n)t}\\
	\notag
	&\times\int_0^te^{C(\lambda_m+\lambda_n)\tau}\delta_{mn}d\tau \\
	&= \frac{D^2\lambda_m^{1/4}\lambda_n^{1/4}(1-e^{-C(\lambda_m+\lambda_n)t})}{L_xC(\lambda_m+\lambda_n)}\delta_{mn},
\end{align}
as expected.

In this subsection we (i) derived the amplitude of the wave modes $\langle c_n^2 \rangle$ directly from the STFE as a function of time, which confirms the result from thermal-capillary-wave theory Eq.~(\ref{eq:equipartitionPinned}) at thermal equilibrium (as $t\to\infty$ and (ii) showed that the wave modes are uncorrelated, which is used in the derivation Eq.~(\ref{eq:2d_pinned_dh22}) of the fluctuation amplitude $\langle h_2^2\rangle$.

\subsection{Solving the linearized problem with Langevin motions on the boundaries}
\label{app:SolveLangevin}
The problem we are looking to solve is given by
\begin{align}
\label{eq:linearLangevin}
    &\frac{\partial h_3}{\partial t} = -\frac{\gamma h_0^3}{3\mu}\frac{\partial^4 h_3}{\partial x^4},\\
    &h_3(0,t) = N_1(t),\: h_3(L_x,t) = N_2(t),\\
    &\frac{\partial^3 h_3}{\partial x^3}(0,t) = \frac{\partial^3 h_3}{\partial x^3}(L_x,t) = 0,
\end{align}
where $N_1(t)$ and $N_2(t)$ are Langevin diffusion process described as
\begin{align}
\label{eq:appELangevin}
	\xi\frac{dN_1}{dt} &= -kN_1(t) + f_1(t),\\
	\xi\frac{dN_2}{dt} &= -kN_2(t) + f_2(t).
\end{align}
Here $\xi$ is the coefficient of friction, $k$ is the harmonic constant. $f_1(t)$ and $f_2(t)$ are Gaussian noises that satisfy $\langle f_1(s)f_1(\tau)\rangle=2\xi k_B T \delta(s-\tau)$ and $\langle f_2(s)f_2(\tau)\rangle=2\xi k_B T \delta(s-\tau)$.
This problem can be further divided into two sub-problems
\begin{subequations}
\begin{align}
    &\frac{\partial h_{31}}{\partial t} = -\frac{\gamma h_0^3}{3\mu}\frac{\partial^4 h_{31}}{\partial x^4},\\
    &h_{31}(0,t) = N_1(t),\: h_{31}(L_x,t) = 0,\\
    &\frac{\partial^3 h_{31}}{\partial x^3}(0,t) = \frac{\partial^3 h_{31}}{\partial x^3}(L_x,t) = 0,
\end{align}
\end{subequations}
and
\begin{subequations}
\begin{align}
    &\frac{\partial h_{32}}{\partial t} = -\frac{\gamma h_0^3}{3\mu}\frac{\partial^4 h_{32}}{\partial x^4},\\
    &h_{32}(0,t) = 0,\: h_{32}(L_x,t) = N_2(t),\\
    &\frac{\partial^3 h_{32}}{\partial x^3}(0,t) = \frac{\partial^3 h_{32}}{\partial x^3}(L_x,t) = 0.
\end{align}
\end{subequations}
Since Eq.~(\ref{eq:linearLangevin}) is linear it is easy to see that $h_3(x,t) = h_{31}(x,t)+h_{32}(x,t)$ is a solution. Now, $h_{31}(x,t)$ can be found with the following procedure. Let $h_{31}(x,t) = w_1(x,t) + v_1(x,t)$, where $w_1(x,t) = N_1(t)(1-x/L_x)^2$. This choice of $w_1(x,t)$ ensures that
\begin{equation}
    w_1(0,t) = N_1(t),\: w_1(L_x,t) = 0,
\end{equation}
and 
\begin{equation}
    \frac{\partial^3 w_1}{\partial x^3} = 0.
\end{equation}
So after substituting $h_{31}(x,t) = w_1(x,t)+v_1(x,t)$ and Langevin diffusion to Eq.~(\ref{eq:linearLangevin}), we have
\begin{equation}
\label{eq:appEvw}
\frac{\partial v_1}{\partial t} + \frac{\partial w_1}{\partial t}= -\frac{\gamma}{3\mu}h_0^3\left(\frac{\partial^4 v_1}{\partial x^4}\right)
\end{equation}
\begin{equation}
v_1(0,t) = v_1(L_x,t) = 0,
\end{equation}
\begin{equation}
\frac{\partial^3 v_1}{\partial x^3}(0,t) = \frac{\partial^3 v_1}{\partial x^3}(L,t) = 0,
\end{equation}
\begin{equation}
N_1(0)\left(1-\frac{x}{L_x}\right)^2 + v_1(x,0) = 0.
\end{equation}
We then expand $v_1(x,t)$ and $\partial w_1(x,t)/\partial t$ in wave modes $\varphi_n(x)$ corresponding to the pinned contact line problem since it is a Dirichlet type boundary condition
\begin{equation}
v_1(x,t) = \alpha_{10}(t)\varphi_0(x) + \sum_{i=1}^{\infty}\alpha_{1i}(t)\varphi_i(x),
\end{equation}
\begin{equation}
\frac{\partial w}{\partial t}(x,t) = \beta_{10}(t)\varphi_0(x) + \sum_{i=1}^{\infty}\beta_{1i}(t)\varphi_i(x).
\end{equation}
Since the expression for $\frac{\partial w}{\partial t}$ is known, we should be able to calculate $\beta_{1i}(t)$ explicitly. However, since $\varphi_i(x)$ are not orthogonal we can only expand it with finitely many wave modes and calculate $\beta_{1i}(t)$ by solving the linear system of finite unknowns. So
\begin{equation}
\label{eq:appEv}
v_1(x,t) = \sum_{i=1}^{N}\alpha_{1i}(t)\varphi_i(x),
\end{equation}
\begin{equation}
\label{eq:appEw}
\frac{\partial w}{\partial t}(x,t) = \sum_{i=1}^{N}\beta_{1i}(t)\varphi_i(x).
\end{equation}
We then multiply both sides of Eq.~(\ref{eq:appEw}) with $\varphi_i(x)$ and integrate w.r.t. $x$ from $[0,L_x]$ to get
\begin{align}
	&\int_0^{L_x}\frac{\partial w}{\partial t}(x,t)\varphi_i(x)dx \\
	&= \int_0^{L_x} \frac{d N_1}{dt}(t)(1-\frac{x}{L})^2\varphi_i(x)dx \\
	&= 
	\begin{cases}
	\frac{dN_1}{dt}(t)\frac{L_x^2}{20},\;\text{for $i=0$}\\
	-\frac{dN_1}{dt}(t)\frac{4}{L_x\lambda_i^{1/2}}\tan\left(\frac{L_x\lambda_i^{1/4}}{2}\right),\;\text{for odd $i$}\\
	\frac{dN_1}{dt}(t)\frac{4}{L_x^2\lambda_i^{3/4}}\left[-2+L_x\lambda_i^{1/4}\cot\left(\frac{L_x\lambda_i^{1/4}}{2}\right)\right],\;\text{for even $i$}
	\end{cases} \\
	&=\sum_{j=0}^{N}\beta_{1j}(t)\int_0^{L_x}\phi_j(x)\phi_i(x)dx\\
	&= \sum_{j=0}^{N}\beta_{1j}(t)G_{ij},
\end{align}
where $G$ is the matrix with
\begin{widetext}
\begin{align}
	G_{ij} &= \int_0^{L_x}\varphi_i(x)\varphi_j(x)dx \\
	&=
	\begin{cases}
	\frac{L_x^3}{30},\;\text{for $i=j=0$}\\
	4\frac{2-L_x\lambda_i^{1/4}\cot(L_x\lambda_i^{1/4}/2)}{L_x\lambda_i^{3/4}},\;\text{for $i=0$, $j$ is even or $j=0$, $i$ is even}\\
	\frac{\tan(\lambda_i^{1/4}L_x/2)(\lambda_i^{1/4}L_x\tan(\lambda_i^{1/4}L_x/2)+2)}{\lambda_i^{1/4}},\;\text{for $i=j$ and $i$ is odd},\\
	\frac{\tan(\lambda_i^{1/4}L_x/2)(\lambda_i^{1/4}L_x\cot(\lambda_i^{1/4}L_x/2)-2)}{\lambda_i^{1/4}},\;\text{for $i=j$ and $i$ is even},\\
	\frac{8\lambda_i^{1/4}\lambda_j^{1/4}(\lambda_i^{1/4}\tan(\lambda_i^{1/4}L_x/2)-\lambda_j^{1/4}\tan(\lambda_j^{1/4}L_x/2))}{\lambda_i-\lambda_j},\;\text{for $i\neq j$ and both odd}\\
	\frac{8\lambda_i^{1/4}\lambda_j^{1/4}(\lambda_j^{1/4}\cot(\lambda_j^{1/4}L_x/2)-\lambda_i^{1/4}\cot(\lambda_i^{1/4}L_x/2))}{\lambda_i-\lambda_j},\;\text{for $i\neq j$ and both even}\\
	0,\;\text{otherwise}.
	\end{cases}
\end{align}
\end{widetext}
Then $\beta_{1i}(t)$ can be solved explicitly in the form of $u_{1i}$
\begin{equation}
\label{eq:appEui}
	\beta_{1i}(t) = \frac{dN_1}{dt}(t)u_{1i}.
\end{equation}
Substituting in Eq.~(\ref{eq:appELangevin}) we have
\begin{equation}
\label{eq:appEbeta}
	\beta_{1i}(t) = u_{1i}(-\frac{k}{\xi}N_1(t)+\frac{1}{\xi}f_1(t)).
\end{equation}
Now substitute Eq.~(\ref{eq:appEv}), Eq.~(\ref{eq:appEw}) and Eq.~(\ref{eq:appEbeta}) into Eq.~(\ref{eq:appEvw}) we get
\begin{align}
\notag
	&\sum_{i=0}^{N}\left(\frac{d\alpha_{1i}}{dt}(t)\varphi_i(x)+(-\frac{k}{\xi}N_1(t)+\frac{1}{\xi}f_1(t))u_{1i}\varphi_i(x)\right) \\
	&= \sum_{i=1}^{N}-\frac{\gamma}{3\mu}h_0^3\alpha_{1i}(t)\frac{d^4\varphi_i}{dx^4}(x).
\end{align}
Recalling
\begin{equation}
	\frac{d^4\varphi_i}{dx^4}(x) = \lambda_i \varphi_i(x),
\end{equation}
we have for $\lambda_0=0$
\begin{equation}
	\left(\frac{d \alpha_{10}}{dt}(t)+u_{10}\frac{dN_1}{dt}(t)\right)\varphi_0(x)=0.
\end{equation}
and for $i=1,2,\ldots,N$
\begin{equation}
	\frac{d\alpha_{1i}}{dt}+\frac{\gamma}{3\mu}h_0^3\lambda_i\alpha_{1i} = (\frac{k}{\xi}N_1(t)-\frac{1}{\xi}f_1(t))u_{1i}.
\end{equation}
Denote $C = \frac{\gamma h_0^3}{3\mu}$ and multiply both side with $\exp(C\lambda_i t)$ we have
\begin{align}
\notag
	\frac{d}{dt}&(\exp(C\lambda_i t)\alpha_{1i}(t)) \\
	&= u_{1i}(\frac{k}{\xi}N_1(t)-\frac{1}{\xi}f_1(t))\exp(C\lambda_i t),
\end{align}
integrate both side we have
\begin{align}
\notag
	\exp(C\lambda_it)&\alpha_{1i}(t)-\alpha_{1i}(0) \\
	&= u_{1i}\int_0^t(\frac{k}{\xi}N_1(\tau)-\frac{1}{\xi}f_1(\tau))\exp(C\lambda_i \tau)d\tau.
\end{align}
Since the initial shape of the free surface is flat we know $\alpha_{1i}(0) = 0$. Then for $i=0$ we have
\begin{equation}
	\alpha_{10}(t) = -u_{10}N_1(t),
\end{equation}
and for $i=1,2,\ldots,N$
\begin{align}
\notag
	\alpha_{1i}(t) &= u_{1i}\exp(-C\lambda_i t)\times\\
	\label{eq:appEalphai}
	&\int_0^t(\frac{k}{\xi}N_1(\tau)-\frac{1}{\xi}f_1(\tau))\exp(C\lambda_i \tau)d\tau.
\end{align}

Thus we have an expression for $h_{31}(x,t)$
\begin{align}
\notag
	h_{31}(x,t) &= \sum_{i=1}^{N}\alpha_{1i}(t)\varphi_i(x)\\
	\label{eq:appEh31}
	&+ N_1(t)\left(1-\frac{x}{L_x}\right)\left(1-\frac{x}{L_x}-u_{10}\frac{x}{L_x}\right).
\end{align}
With the same procedure we can solve for $h_{32}$ as well
\begin{align}
\notag
	h_{32}(x,t) &= \sum_{i=1}^{N}\alpha_{2i}(t)\varphi_i(x)\\
	\label{eq:appEh32}
	&+ N_2(t)\frac{x}{L_x}\left(\frac{x}{L_x}-u_{20}(1-\frac{x}{L_x})\right),
\end{align}
where for $i=1,2,\ldots,N$
\begin{align}
\notag
	\alpha_{2i}(t) &= u_{2i}\exp(-C\lambda_i t)\times\\
	&\int_0^t(\frac{k}{\xi}N_2(\tau)-\frac{1}{\xi}f_2(\tau))\exp(C\lambda_i \tau)d\tau,
\end{align}
and $u_{2i}$ is obtained similar to Eq.~(\ref{eq:appEui}). And we can write out $h_3(x,t)$ in the required form
\begin{align}
\notag
    h_3(x,t) &= \sum_{i=1}^N e_i(t)\varphi_i(x)\\
    \notag
    &+ N_1(t)\left(1-\frac{x}{L_x}\right)\left(1-\frac{x}{L_x}-u_{10}\frac{x}{L_x}\right)\\
    &+ N_2(t)\frac{x}{L_x}\left(\frac{x}{L_x}-u_{20}(1-\frac{x}{L_x})\right),
\end{align}
where $e_{i}(t) = \alpha_{1i}(t)+\alpha_{2i}(t)$.

In this subsection we solved the linearized TFE with Langevin diffusion motion on the boundaries analytically and give the details of the derivation of Eq.~(\ref{eq:2d_pinned_h3}).

\subsection{Combined fluctuation amplitude}

\label{app:Langevin_dh2_SLE}

We first show that $\langle h_2h_3\rangle=0$. From Appendix \ref{app:SolveLangevin} we know $h_3(x,t) = h_{31}(x,t) + h_{32}(x,t)$, so
\begin{equation}
    \langle h_2h_3\rangle = \langle h_2h_{31}\rangle + \langle h_2h_{32}\rangle.
\end{equation}
By Eq.~(\ref{eq:pinned_h2}) and Eq.~(\ref{eq:appEh31}) we have
\begin{align}
\notag
    \langle h_2h_{31}\rangle &= \sum_{i=1}^{\infty}\sum_{j=1}^{N}\langle c_i(t)\alpha_{1j}(t)\rangle \varphi_i(x)\varphi_j(x)\\
    \notag
        &+\sum_{i=1}^{\infty}\langle N_1(t)c_i(t)\rangle\varphi_i(x)\left(1-\frac{x}{L_x}\right)\\
        &\times\left(1-\frac{x}{L_x}-u_{10}\frac{x}{L_x}\right).
\end{align}
By Eq.~(\ref{eq:appDcn}) and Eq.~(\ref{eq:appEalphai}) we have
\begin{align}
\label{eq:appFcalpha}
\notag
    \langle c_i(t)\alpha_{1j}(t)\rangle &=Du_{1j}\lambda_i^{1/4}e^{-C(\lambda_i+\lambda_j) t}\\
    \notag
    &\times \int_0^t\int_0^t e^{C(\lambda_i\tau+\lambda_j s)}\frac{1}{\xi}(k\langle N_1(s)d_i(\tau)\rangle\\
    &-\langle f_1(s)d_i(\tau)\rangle) d\tau ds
\end{align}
where $N_1(s)$ is the position of the fluctuating contact-line driven by random force $f_1(s)$ and $d_i(\tau)$ is random flux.
By Eq.~(\ref{eq:appDcn}) we have
\begin{align}
\label{eq:appFNc}
    \langle N_1(t)c_i(t)\rangle &= D\lambda_i^{1/4}e^{-C\lambda_i t}\int_0^te^{C\lambda_i \tau}\langle d_i(\tau)N_1(t)\rangle d\tau.
\end{align}
If we consider that the random force $f_1(s)$ is uncorrelated to random flux $d_i(\tau)$, then by Eq.~(\ref{eq:appFcalpha}) we have $\langle c_i(t)a_{1j}(t)\rangle = 0$ and by Eq.~(\ref{eq:appFNc}) we have $\langle N_1(t) c_i(t)\rangle = 0$ and thus $\langle h_2 h_{31}\rangle = 0$.
Applying the same argument one can derive that $\langle h_2 h_{32}\rangle = 0$, and thus $\langle h_2h_3\rangle=0$.

Now we consider $\langle h_3^2\rangle = \langle h_{31}^2\rangle+2\langle h_{31}h_{32}\rangle+\langle h_{32}^2\rangle$. We first show that $\langle h_{31}h_{32}\rangle=0$. If we assume that the random forces driving the fluctuations of the contact-lines are uncorrelated $\langle f_1(t)f_2(t)\rangle=0$ then the positions of the contact-lines are also uncorrelated $\langle N_1(t)N_2(t)\rangle=0$. By Eq.~(\ref{eq:appEh31}) and Eq.~(\ref{eq:appEh32}) we have
\begin{align}
\notag
    \langle h_{31}&h_{32}\rangle = \sum_{i=1}^{N}\sum_{j=1}^N\langle \alpha_{1i}(t)\alpha_{2j}(t)\rangle\varphi_i(x)\varphi_j(x)\\
    \notag
    &+\sum_{i=1}^N\langle \alpha_{1i}(t)N_2(t)\rangle \varphi_i(x)\frac{x}{L_x}\left(\frac{x}{L_x}-u_{20}(1-\frac{x}{L_x})\right)\\
    \notag
    &+\sum_{j=1}^N\langle \alpha_{2j}(t)N_1(t)\rangle \left(1-\frac{x}{L_x}\right)\left(1-\frac{x}{L_x}-u_{10}\frac{x}{L_x}\right)\\
    \notag
    &+\langle N_1(t)N_2(t)\rangle\left(1-\frac{x}{L_x}\right)\left(1-\frac{x}{L_x}-u_{10}\frac{x}{L_x}\right)\\
    &\times\frac{x}{L_x}\left(\frac{x}{L_x}-u_{20}(1-\frac{x}{L_x})\right)\\
    &=0
\end{align}
We then consider $\langle h_{31}^2\rangle$. By Eq.~(\ref{eq:appEh31}) we can calculate
\begin{align}
\notag
    \langle h_{31}^2\rangle &= \sum_{i=1}^N\sum_{j=1}^N\langle \alpha_{1i}(t)\alpha_{1j}(t)\rangle \varphi_i(x)\varphi_j(x)\\
    \notag
    &+2\sum_{i=1}^N \langle \alpha_{1i}(t)N_1(t)\rangle\varphi_i(x)\left(1-\frac{x}{L_x}\right)\\
    \notag
    &\times\left(1-\frac{x}{L_x}-u_{10}\frac{x}{L_x}\right)\\
    &+\langle N_1^2(t)\rangle \left(1-\frac{x}{L_x}\right)^2\left(1-\frac{x}{L_x}-u_{10}\frac{x}{L_x}\right)^2.
\end{align}
By Eq.~(\ref{eq:appEalphai}) we have
\begin{align}
\notag
    &\langle \alpha_{1i}(t)\alpha_{1j}(t)\rangle = \langle\frac{u_{1i}u_{1j}k^2}{\xi^2}e^{-A(\sigma_i+\sigma_j)t}\\
    \notag
    &\times\int_0^t\int_0^tN_1(\tau)N_1(s)e^{A(\sigma_i\tau+\sigma_js)}d\tau ds\rangle\\
    \notag
    &-\langle \frac{2u_{1i}u_{1j}k}{\xi^2}e^{-A(\sigma_i+\sigma_j)t}\\
    \notag
    &\times\int_0^t\int_0^tN_1(\tau)f(s)e^{A(\sigma_i\tau+\sigma_js)}d\tau ds\rangle\\
    \notag
    &+\langle\frac{u_{1i}u_{1j}}{\xi^2}e^{-A(\sigma_i+\sigma_j)t}\\
    &\times\int_0^t\int_0^tf(\tau)f(s)e^{A(\sigma_i\tau+\sigma_js)}d\tau ds\rangle
\end{align}
and
\begin{align}
\notag
    &2\langle \alpha_{1i}(t)N_1(t)\rangle = \langle\frac{2u_{1i}k}{\xi}e^{-C\sigma_i t}\\
    \notag
    &\times\int_0^tN_1(\tau)N_1(t)e^{C\sigma_i\tau}d\tau \rangle\\
    &-\langle\frac{2u_{1i}}{\xi}e^{-C\sigma_i t}\int_0^tf(\tau)N_1(t)e^{C\sigma_i\tau}d\tau\rangle
\end{align}
It is well known that for Langevin process
\begin{equation}
\label{eq:appLangevinCOV}
	\langle N_1(\tau)N_1(s)\rangle = \frac{k_B T}{k}e^{-\frac{k}{\xi}|\tau-s|},
\end{equation}
and
\begin{equation}
	\langle f(\tau)f(s)\rangle = 2\xi k_B T \delta(\tau-s).
\end{equation}
But we don't know what $\langle N_1(\tau)f(s)\rangle$ is. By Langevin diffusion equation we know
\begin{equation}
\frac{dN_1}{dt} = -\frac{k}{\xi}N_1(t) + \frac{1}{\xi}f(t),
\end{equation}
so
\begin{align}
\frac{d}{dt}\left(e^{k/\xi t}N_1(t)\right) = \frac{1}{\xi}e^{k/\xi t}f(t),
\end{align}
then
\begin{equation}
e^{k/\xi t}N_1(t)-N_1(0) = \frac{1}{\xi}\int_0^{t}e^{k/\xi\tau}f(\tau)d\tau.
\end{equation}
If we let $N_1(0) = 0$, we have
\begin{equation}
N_1(t) = \frac{1}{\xi}e^{-k/\xi t}\int_0^{t}e^{k/\xi\tau}f(\tau)d\tau.
\end{equation}
Then we have
\begin{align}
\notag
\langle N_1(\tau)f(s)\rangle &= \langle \frac{1}{\xi}e^{-k/\xi\tau}\int_0^{\tau}e^{k/\xi r}f(r)dr f(s)\rangle \\
\notag
&= \frac{1}{\xi}e^{-k/\xi\tau}\int_0^{\tau}e^{k/\xi r}\langle f(r)f(s)\rangle dr\\
\notag
&= 2k_B Te^{-k/\xi \tau}\int_0^{\tau}e^{k/\xi r}\delta(r-s)dr\\
&=
\begin{cases}
2k_BTe^{-k/\xi(\tau-s)},\;\text{if } \tau>s \\
0,\;\text{if } \tau\leq s
\end{cases}
\end{align}
So with careful evaluation we come to
\begin{equation}
    \langle h_{31}^2(x,t)\rangle = \sum_{k=1}^9 \mathcal{S}_k(x,t),
\end{equation}
where
\begin{align}
\notag
    \mathcal{S}_1(x,t) &= \sum_{i=1}^N\sum_{j=1}^N\frac{k_BT u_{1i} u_{1j} k}{\xi^2}\phi_i(x)\phi_j(x)\\
	&\times\left(\frac{1-\exp(-C(\sigma_i+\sigma_j)t)}{(C\sigma_i+\frac{k}{\xi})(C(\sigma_i+\sigma_j))}\right),
\end{align}
\begin{align}
\notag
    \mathcal{S}_2(x,t) &= -\sum_{i=1}^N\sum_{j=1}^N\frac{k_BT u_{1i} u_{1j} k}{\xi^2}\phi_i(x)\phi_j(x)\\
	&\times\left(\frac{\exp(-(C\sigma_i+\frac{k}{\xi})t)-\exp(-C(\sigma_i+\sigma_j)t)}{(C\sigma_i+\frac{k}{\xi})(C\sigma_j-\frac{k}{\xi})}\right),
\end{align}
\begin{align}
\notag
    \mathcal{S}_3(x,t) &=\sum_{i=1}^N\sum_{j=1}^N\frac{k_BT u_{1i} u_{1j} k}{\xi^2}\phi_i(x)\phi_j(x)\\
	&\times\left(\frac{1-\exp(-(C\sigma_j+\frac{k}{\xi})t)}{(C\sigma_i-\frac{k}{\xi})(C\sigma_j+\frac{k}{\xi})}\right),
\end{align}
\begin{align}
\notag
    \mathcal{S}_4(x,t) &=-\sum_{i=1}^N\sum_{j=1}^N\frac{k_BT u_{1i} u_{1j} k}{\xi^2}\phi_i(x)\phi_j(x)\\
	&\times\left(\frac{1-\exp(-C(\sigma_i+\sigma_j)t)}{(C\sigma_i-\frac{k}{\xi})(C(\sigma_i+\sigma_j))}\right),
\end{align}
\begin{align}
\notag
    \mathcal{S}_5(x,t) &=\sum_{i=1}^N\sum_{j=1}^N\frac{2 k_B Tu_{1i} u_{1j}}{\xi}\phi_i(x)\phi_j(x) \\
	&\times\left(\frac{1-\exp(-C(\sigma_i+\sigma_j)t)}{(C(\sigma_i+\sigma_j))}\right) ,
\end{align}
\begin{align}
    \mathcal{S}_6(x,t) &=\frac{k_B T}{k}\left(1-\frac{x}{L}\right)^2\left(1-\frac{x}{L}-u_0x\right)^2,
\end{align}
\begin{align}
\notag
    \mathcal{S}_7(x,t) &=-\sum_{i=1}^N\sum_{j=1}^N\frac{4k_B Tu_{1i} u_{1j} k}{\xi^2}\phi_i(x)\phi_j(x)\\
    &\times\left(\frac{1-\exp(-(C\sigma_j+\frac{k}{\xi})t)}{(C\sigma_i-\frac{k}{\xi})(C\sigma_j+\frac{k}{\xi})}\right),
\end{align}
\begin{align}
\notag
    \mathcal{S}_8(x,t) &=+\sum_{i=1}^N\sum_{j=1}^N\frac{4k_B Tu_{1i}u_{1j}k}{\xi^2}\phi_i(x)\phi_j(x)\\
    &\times\left(\frac{1-\exp(-C(\sigma_i+\sigma_j)t)}{(C\sigma_i-\frac{k}{\xi})(C(\sigma_i+\sigma_j))}\right),
\end{align}
\begin{align}
\notag
    \mathcal{S}_9(x,t) &=-\sum_{i=1}^N\left(\frac{1-\exp(-(C\sigma_i+\frac{k}{\xi})t)}{(C\sigma_i+\frac{k}{\xi})}\right)\\
    &\times\frac{2k_BTu_{1i}}{\xi}\phi_i(x)\left(1-\frac{x}{L}\right)\left(1-\frac{x}{L}-u_0x\right).
\end{align}
$k$ and $\xi$ can be extracted from MD simulations via Eq.~(\ref{eq:appLangevinCOV}), and we found that $C\sigma_i+k/\xi$ and $C(\sigma_i+\sigma_j)$ are always positive for any $i$ and $j$, so as $t\to\infty$, $\mathcal{S}_2=0$, $\mathcal{S}_3$ merges with $\mathcal{S}_7$, $\mathcal{S}_4$ merges with $\mathcal{S}_8$, and we have
\begin{align}
\label{eq:apph31FA}
\notag
\langle h_{31}^2(x)\rangle &= \sum_{i=1}^N\sum_{j=1}^N\frac{k_BT u_{1i}u_{1j} k}{\xi^2(A\sigma_i+\frac{k}{\xi})(A(\sigma_i+\sigma_j))}\phi_i(x)\phi_j(x)\\
\notag
&+ \sum_{i=1}^N\sum_{j=1}^N\frac{2 k_B Tu_{1i}u_{1j}}{\xi(A(\sigma_i+\sigma_j))} \phi_i(x)\phi_j(x) \\
\notag
&+\frac{k_B T}{k}\left(1-\frac{x}{L}\right)^2\left(1-\frac{x}{L}-u_0x\right)^2\\
\notag
&-\sum_{i=1}^N\sum_{j=1}^N\frac{3k_B Tu_{1i}u_{1j}k}{\xi^2(A\sigma_i-\frac{k}{\xi})(A\sigma_j+\frac{k}{\xi})}\phi_i(x)\phi_j(x)\\
\notag
&+\sum_{i=1}^N\sum_{j=1}^N\frac{3k_B Tu_{1i}u_{1j}k}{\xi^2(A\sigma_i-\frac{k}{\xi})(A(\sigma_i+\sigma_j))}\phi_i(x)\phi_j(x)\\
\notag
&-\sum_{i=1}^N\frac{2k_BTu_{1i}}{\xi(A\sigma_i+\frac{k}{\xi})}\phi_i(x)\\
&\times\left(1-\frac{x}{L}\right)\left(1-\frac{x}{L}-u_0x\right).
\end{align}
Following the same derivation one can find that $\langle h_{32}^2(x)\rangle$ is symmetric to $\langle h_{31}^2(x)\rangle$ around $L_x/2$. And so with Eq.~(\ref{eq:apph31FA}) we have calculated 
\begin{equation}
\label{eq:apph3FA}
    \langle h_3^2(x)\rangle = \langle h_{31}^2(x)\rangle + \langle h_{32}^2(x)\rangle.
\end{equation} 

In this subsection we (i) showed that perturbation induced by thermal noise in the bulk $h_2$ and perturbation induced by thermal noise on the boundary $h_3$ are uncorrelated under the linearized construction and (ii) calculated the fluctuation amplitude $\langle h_3^2(x)\rangle$ and the combined fluctuation amplitude $\langle (h_2(x)+h_3(x))^2\rangle$, expression Eq.~(\ref{eq:TCW_pin}) in details.

\section{3D circular thin film with $90^\circ$ contact angle}
In this section we layout the technical details for 3D circular thin film with $90^\circ$ contact angle. 
\subsection{Derivation of wave modes }
\label{app:3dmodes_90}
We begin with the linearization of thin-film equation in 3D. Consider a perturbation to the free surface 
\begin{equation}
    h(\bm{x},t) = h_0 + \epsilon h_4(\bm{x})T(t),
\end{equation}
where $\epsilon\ll 1$. Apply the perturbation to 3D TFE Eq.~(\ref{eq:3dLE}) and match the leading order we get
\begin{equation}
    \frac{d T}{dt}h_1(\bm{x}) = -\frac{\gamma h_0^3}{3\mu}\nabla^4 h_4(\bm{x}),
\end{equation}
where $\nabla^4$ is the biharmonic operator. We can then separate the variables to get
\begin{equation}
    \frac{T'}{T} = -\frac{\gamma h_0^3}{3\mu}\frac{\nabla^2\nabla^2h_4}{h_4} = -\lambda^4,
\end{equation}
where the minus sign in front of $\lambda^4$ guarantees stability, given $\lambda$ being real number. Since we have a circular thin film it is natural to work in cylindrical coordinate and we arrive at the following eigenvalue problem
\begin{equation}
    \nabla^2\nabla^2 h_4(r,\theta) = \omega^4h_4(r,\theta)
\end{equation}
where $\omega^4 = \lambda^4(3\mu)/(\gamma h_0^3)>0$. The general solution of the eigenvalue problem of the biharmonic operator can be obtained in cylindrical coordinate as follows. For the moment let us denote $h_4$ as $H$ and the eigenvalue problem can be rewritten as
\begin{equation}
\label{eq:app3deigen}
    (\nabla^2-\omega^2)(\nabla^2+\omega^2)H(r,\theta)=0.
\end{equation}
This tells us that $H = C_1 H_1 + C_2 H_2$ where
\begin{equation}
	\nabla^2 H_1 + \omega^2 H_1 = 0
\end{equation}
and
\begin{equation}
	\nabla^2 H_2 - \omega^2 H_2 = 0,
\end{equation}
and $C_1$ $C_2$ are constants.
This is easy to see: Eq.~(\ref{eq:app3deigen}) is equivalent to 
\begin{equation}
\begin{cases}
\nabla^2H_1(r,\theta) + \omega^2H_1(r,\theta) = 0,\\
\nabla^2H(r,\theta) - \omega^2H(r,\theta) = H_1(r,\theta),
\end{cases}
\end{equation}
or
\begin{equation}
	\begin{cases}
	\nabla^2H_2(r,\theta) - \omega^2H_2(r,\theta) = 0,\\
	\nabla^2H(r,\theta) + \omega^2H(r,\theta) = H_2(r,\theta).
	\end{cases}
\end{equation}
For sake of argument, continue with the first case (and later it is easy to see that the two cases are equivalent) - we have already worked out what $H_1$ is. From the equation for $H$ we can see that $H$ is the solution of the homogeneous problem plus a special solution, and it is easy to see that
\begin{equation}
	H = H_2 - \frac{1}{2\omega^2}H_1.
\end{equation}
To solve for $H_1$ and $H_2$ we use separation of variables $H_1(r,\theta) = R_1(r)\Theta_1(\theta)$, $H_2(r,\theta) =  R_2(r)\Theta_2(\theta)$, and we have
\begin{equation}
R_1''\Theta_1 + \frac{1}{r}R_1'\Theta_1 +\frac{1}{r^2}R_1\Theta_1''+\omega^2R_1\Theta_1 = 0.	
\end{equation}
\begin{equation}
	R_2''\Theta_2 + \frac{1}{r}R_2'\Theta_2 +\frac{1}{r^2}R_2\Theta_2''-\omega^2R_2\Theta_2 = 0.	
\end{equation}
Divide by $R_1\Theta_1$ ($R_2\Theta_2$) we have
\begin{equation}
-\frac{\Theta_1''}{\Theta_1} = r^2\frac{R_1''}{R_1}+r\frac{R_1'}{R_1}+\omega^2r^2 = n^2.
\end{equation}
\begin{equation}
	-\frac{\Theta_2''}{\Theta_2} = r^2\frac{R_2''}{R_2}+r\frac{R_2'}{R_2}-\omega^2r^2 = n^2.
\end{equation}
Since $\Theta_1$ and $\Theta_2$ must be periodic with $2\pi$, we have
\begin{equation}
	\Theta_1(\theta) = A_1\cos(n\theta)+B_1\sin(n\theta),
\end{equation}
\begin{equation}
\Theta_2(\theta) = A_2\cos(n\theta)+B_2\sin(n\theta),
\end{equation}
where $n$ is a positive integer. And for $R_1$, $R_2$, if we let $\rho = \omega r$ we have
\begin{equation}
\label{eq:bessel1}
R_1'' + \frac{1}{\rho}R_1' + \left(1-\frac{n^2}{\rho^2}\right)R_1 = 0,
\end{equation}
\begin{equation}
\label{eq:bessel2}
	R_2'' + \frac{1}{\rho}R_2' - \left(1+\frac{n^2}{\rho^2}\right)R_2 = 0,
\end{equation}
and we can immediately see that these are the Bessel function of the first kind and the modified Bessel function of the first kind, so
\begin{equation}
	R_1(\rho) = C_1 J_n(\rho) + D_1 Y_n(\rho),
\end{equation} 
\begin{equation}
	R_2(\rho) = C_2 I_n(\rho) + D_2 K_n(\rho).
\end{equation}
And so
\begin{align}
\notag
	H_n(r,\theta) &= (A_1\cos(n\theta)+B_1\sin(n\theta))\\
	\notag
	&\times(C_1 J_n(\omega r) + D_1 Y_n(\omega r))\\
	\notag
	&+(A_2\cos(n\theta)+B_2\sin(n\theta))\\
	&\times(C_2 I_n(\omega r) + D_2 K_n(\omega r)).
\end{align}
Since height of the film at the origin is finite, the terms involving $Y_n$ and $K_n$ must be zero, so the general solution is
\begin{align}
\label{eq:3dGeneralSoln}
\notag
	H_n(r,\theta) &= (A_1\cos(n\theta)+B_1\sin(n\theta))J_n(\omega r) \\
	&+ (A_2\cos(n\theta)+B_2\sin(n\theta))I_n(\omega r),
\end{align}
where $n=0,1,\ldots$. Here we don't have to consider negative $n$ because
\begin{equation}
	J_{-n} = (-1)^nJ_n
\end{equation}
and
\begin{equation}
	I_{-n} = I_n.
\end{equation}
Applying the $90^\circ$-contact-angle boundary condition Eq.~(\ref{eq:3dBC90}) we obtain
\begin{equation}
\Theta_1 J_n'(\omega a) + \Theta_2 I_n'(\omega a) = 0,
\end{equation}
and applying the no-flux boundary condition Eq.~(\ref{eq:3dBCnoflux}) gives
\begin{equation}
    -\Theta_1 J_n'(\omega a)+\Theta_2 I_n'(\omega a)=0.
\end{equation}
These two conditions tell us the dispersion relation
\begin{equation}
    J_n'(\omega a) = 0,
\end{equation}
and since $I_n'$ is always positive,
\begin{equation}
    \Theta_2=0.
\end{equation}
For each $n$, the dispersion relation gives a list of suitable frequencies $\{ \omega_{n,\alpha}:\: \alpha=1,2,\ldots \}$, and so the wave modes are
\begin{align}
    \Upsilon^1_{n,\alpha}(r,\theta) &= \cos(n\theta)\chi_{n,\alpha}(r),\\
    \Upsilon^2_{n,\alpha}(r,\theta) &= \sin(n\theta)\chi_{n,\alpha}(r),
\end{align}
where
\begin{equation}
    \chi_{n,\alpha} = J_n(\omega_{n,\alpha}r),\: n=0,1,\ldots
\end{equation}

In this subsection we (i) derived the general solution to the eigenvalue problem for 3D circular film, which will be used in Eq.~(\ref{eq:app_3d_pin_use_general}) and (ii) derived the wave modes for 3D circular film with prescribed $90^\circ$ contact angle.

\subsection{Thermal-capillary-wave theory}
\label{app:3dTCW}
The free energy required to perturb the free surface is given by the product of the surface tension and the surface area created:
\begin{align}
	E = \gamma\Bigg(\int_0^{2\pi}\int_0^a\sqrt{1+\left(\frac{\partial h}{\partial r}\right)^2+\frac{1}{r^2}\left(\frac{\partial h}{\partial \theta}\right)^2}rdrd\theta-\pi a^2\Bigg)\\
\end{align}
and assuming small perturbations
\begin{equation}
    r^2\left(\frac{\partial h}{\partial r}\right)^2+\left(\frac{\partial h}{\partial \theta}\right)^2\ll 1,
\end{equation}
so that a Taylor expansion gives
\begin{equation}
    E \approx \frac{\gamma}{2}\int_0^{2\pi}\int_0^a\left(\left(\frac{\partial h}{\partial r}\right)^2+\frac{1}{r^2}\left(\frac{\partial h}{\partial \theta}\right)^2\right)rdrd\theta.
\end{equation}
Applying Eq.~(\ref{eq:3D90decomp}) and denoting 
\begin{equation}
    \Theta_{m,\alpha} = A_{m,\alpha}(t)\cos(m\theta)+B_{m,\alpha}\sin(m\theta)
\end{equation}
we find
\begin{align}
\notag
	E&=\frac{\gamma}{2}\sum_{m=0}^{\infty}\sum_{n=0}^{\infty}\sum_{\alpha=1}^{\infty}\sum_{\beta=1}^{\infty}\int_0^{2\pi}\int_0^a\Big[\Theta_{m,\alpha}(\theta)\Theta_{n,\beta}(\theta) \\
	\notag
	&\times\chi'_{m,\alpha}(r)\chi'_{n,\beta}(r)r+\Theta'_{m,\alpha}(\theta)\Theta'_{n,\beta}(\theta)\\
	\notag
	&\times\chi_{m,\alpha}(r)\chi_{n,\beta}(r)\frac{1}{r}\Big]drd\theta\\
	\notag
	&=\gamma\pi\sum_{\alpha=1}^{\infty}\sum_{\beta=1}^{\infty}\int_0^aA_{0,\alpha}A_{0,\beta}\chi'_{0,\alpha}\chi'_{0,\alpha}rdr\\
	\notag
	&+\frac{\gamma\pi}{2}\sum_{m=1}^{\infty} \sum_{\alpha=1}^{\infty}\sum_{\beta=1}^{\infty}\int_0^a(A_{m,\alpha}A_{m,\beta}+B_{m,\alpha}B_{m,\beta})\\
	\notag
	&\times \Big(\chi_{m,\alpha}'\chi_{m,\beta}'r+m^2\chi_{m,\alpha}\chi_{m,\beta}\frac{1}{r}\Big)dr\\
	\notag
	&=\gamma\pi\sum_{\alpha=1}^{\infty}\sum_{\beta=1}^{\infty}A_{0,\alpha}A_{0,\beta}\\
	\notag
	&\times\int_0^a\left(r\chi'_{0,\alpha}\chi'_{0,\beta}+\frac{0^2}{r}\chi_{0,\alpha}\chi_{0,\beta}\right)dr\\
	\notag
	&+\frac{\gamma\pi}{2}\sum_{m=1}^{\infty}\sum_{\alpha=1}^{\infty}\sum_{\beta=1}^{\infty}(A_{m,\alpha}A_{m,\beta}+B_{m,\alpha}B_{m,\beta})\\
	&\times\int_0^a\left(r\chi'_{m,\alpha}\chi'_{m,\beta}+\frac{m^2}{r}\chi_{m,\alpha}\chi_{m,\beta}\right)dr.
\end{align}
We can show that (even for $m=0$)
\begin{align}
\notag
	&\int_0^a\left(r\chi'_{m,\alpha}\chi'_{m,\beta}+\frac{m^2}{r}\chi_{m,\alpha}\chi_{m,\beta}\right)dr \\
	\notag
	=& \int_0^a\left(\frac{m^2}{r}\chi_{m,\alpha}-\chi'_{m,\alpha}-ra\chi''_{m,\alpha}\right)\chi_{m,\beta}dr \\
	\label{eq:appCalcS}
	=& S_{m,\alpha}\delta_{\alpha\beta}
\end{align}
where
\begin{align}
\notag
    S_{m,\alpha}&=\frac{1}{2}\omega_{m,\alpha}a\Bigg(\omega_{m,\alpha}aJ_{m-1}^2(\omega_{m,\alpha}a)\\
    \notag
    &-2mJ_{m-1}(\omega_{m,\alpha}a)J_m(\omega_{m,\alpha}a)\\
    &+\omega_{m,\alpha}aJ_m^2(\omega_{m,\alpha}a)\Bigg).
\end{align}
$S_{m,\alpha}$ can be further simplified using the fact that $\chi_{m,\alpha}=J_m(\omega_{m,\alpha}r)$ and the property of Bessel function $J_m'(\rho)=J_{m-1}-m/\rho J_m(\rho)$
\begin{align}
\notag
	S_{m,\alpha} &= \frac{1}{2}\omega_{m,\alpha}a\Bigg(\omega_{m,\alpha}a(J_m'(\omega_{m,\alpha}a))^2\\
	&-\frac{m^2}{\omega_{m,\alpha}a}J_m^2(\omega_{m,\alpha})+\omega_{m,\alpha}aJ_m^2(\omega_{m,\alpha}a)\Bigg).
\end{align}
The dispersion relation Eq.~(\ref{eq:3d90dispersion}) then tells us that
\begin{equation}
	S_{m,\alpha} = \frac{1}{2}\left(\omega_{m,\alpha}^2a^2-m^2\right)J_m^2(\omega_{m,\alpha}a).
\end{equation}
Considering the asymptotic expansion of the Bessel function of first kind, as $x\to\infty$,
\begin{equation}
\label{eq:appBesselexp}
    J_m(x) \sim \sqrt{\frac{2}{\pi x}}\cos\left(x-\frac{2m+1}{4}\pi\right) + \mathcal{O}(x^{-3/2}),
\end{equation}
one can easily see that since $\omega_{m,\alpha}$ increases with $m$ linearly, $S_{m,\alpha}$ also increases with $m$ linearly for large $m$. Follow the same procedure as in Eq.~(\ref{eq:appCalcS}), we can also show that the wave modes are orthogonal,
\begin{equation}
	\int_0^{2\pi}\int_0^a\Upsilon^i_{m,\alpha}(r,\theta)\Upsilon^j_{n,\beta}(r,\theta) rdrd\theta = \delta_{ij}\delta_{mn}\delta_{\alpha\beta}C,
\end{equation}
for some constant $C$, which gives us confidence to apply equipartition theorem.

In this subsection we (i) calculated the extra surface energy associated with the perturbations supporting Eq.~(\ref{eq:3d_90_energy}), (ii) showed that perturbed surface area $S_{n,\alpha}$ increases linearly with $n$, which leads to the use of cut-off length scale and (iii) showed that the wave modes $\Upsilon^i_{n,\alpha}$ are orthogonal to support Eq.~(\ref{eq:3d_90_equi1}) and Eq.~(\ref{eq:3d_90_equi2}).

\section{3D circular thin film with a pinned contact line}
In this section we layout the technical details for the 3D circular thin film with a pinned contact line.
\subsection{Derivation of wave modes }
\label{app:3dmodes_pinned}
Applying the pinned boundary condition Eq.~(\ref{eq:3dBCpinned}) and no-flux boundary condition Eq.~(\ref{eq:3dBCnoflux}) to the general solution Eq.~(\ref{eq:3dGeneralSoln})
\begin{align}
\notag
	H_n(r,\theta) &= (A_1\cos(n\theta)+B_1\sin(n\theta))\\
	\notag
	&\times(C_1 J_n(\zeta r) + D_1 Y_n(\zeta r))\\
	\notag
	&+(A_2\cos(n\theta)+B_2\sin(n\theta))\\
	\label{eq:app_3d_pin_use_general}
	&\times(C_2 I_n(\zeta r) + D_2 K_n(\zeta r)).
\end{align}
we get
\begin{equation}
    \Theta_1 J_n(\zeta a) + \Theta_2 I_n(\zeta a) = 0
\end{equation}
and
\begin{equation}
    -\Theta_1 J_n'(\zeta a) + \Theta_2 I_n'(\zeta a) = 0.
\end{equation}
This tells us that
\begin{equation}
    \Theta_2 = -\frac{J_n(\zeta a)}{I_n(\zeta a)}\Theta_1
\end{equation}
and gives us the dispersion relation
\begin{align}
    \notag
    2n J_n(\zeta a)I_n(\zeta a) &+ \zeta a\Big[J_n(\zeta a)I_{n+1}(\zeta a)\\
    &-J_{n+1}(\zeta a)I_n(\zeta a)\Big] = 0.
\end{align}
For each $n$ the dispersion relation gives a list of suitable frequencies $\{\zeta_{n,\alpha}:\alpha=1,2,\ldots\}$, with wave modes given by
\begin{align}
    &\Psi^1_{n,\alpha}(r,\theta) = \cos(n\theta)\psi_{n,\alpha}(r),\\
    &\Psi^2_{n,\alpha}(r,\theta) = \sin(n\theta)\psi_{n,\alpha}(r).
\end{align}
Here
\begin{align}
\notag
    \psi_{n,\alpha}(r) = J_n(\zeta_{n,\alpha} r)-\frac{J_n(\zeta_{n,\alpha} a)}{I_n(\zeta_{n,\alpha} a)}&I_n(\zeta_{n,\alpha} r),\\
    & n=0,1,\ldots
\end{align}

\subsection{Thermal-capillary-wave theory}
\label{app:3dpinnedTCW}
Similar to Appendix \ref{app:3dTCW} we have
\begin{align}
    \notag
	S &=\gamma\pi\sum_{\alpha=1}^{\infty}\sum_{\beta=1}^{\infty}C_{0,\alpha}C_{0,\beta}\\
	\notag
	&\times\int_0^a\left(r\psi'_{0,\alpha}\psi'_{0,\beta}+\frac{0^2}{r}\psi_{0,\alpha}\psi_{0,\beta}\right)dr\\
	\notag
	&+\frac{\gamma\pi}{2}\sum_{m=1}^{\infty}\sum_{\alpha=1}^{\infty}\sum_{\beta=1}^{\infty}(C_{m,\alpha}C_{m,\beta}+D_{m,\alpha}D_{m,\beta})\\
	&\times\int_0^a\left(r\psi'_{m,\alpha}\psi'_{m,\beta}+\frac{m^2}{r}\psi_{m,\alpha}\psi_{m,\beta}\right)dr,
\end{align}
where $C_{m,\alpha}$ and $D_{m,\alpha}$ comes from the notation
\begin{equation}
    \Theta_{m,\alpha}(\theta) = C_{m,\alpha}\cos(m\theta) + D_{m,\alpha}\sin(m\theta).
\end{equation}
We now show that for any $m$ (even when $m=0$)
\begin{align}
\notag
	&\int_0^a\left(r\psi'_{m,\alpha}\psi'_{m,\beta}+\frac{m^2}{r}\psi_{m,\alpha}\psi_{m,\beta}\right)dr\\
	=& \int_0^a\left(\frac{m^2}{r}\psi_{m,\alpha}-\psi'_{m,\alpha}-r\psi''_{m,\alpha}\right)\psi_{m,\beta}dr \\
	\notag
	=& K_{m,\alpha}\delta_{\alpha\beta}
\end{align}
where the first equality used the fact that $\psi_{m,\beta}(0)=\psi_{m,\beta}(a)=0$. Recall that $f_{m,\alpha}(r,\theta)=\Theta_{m,\alpha}(\theta)\psi_{m,\alpha}(r)$ is one of the eigenfunctions that satisfies
\begin{equation}
	\nabla^2\nabla^2 f_{m,\alpha}(r,\theta) = \omega_{m,\alpha}^4f_{m,\alpha}(r,\theta),
\end{equation}
and expanding this equation in polar coordinate gives us
\begin{align}
\notag
	&\psi_{m,\alpha}'''' + \frac{2}{r}\psi_{m,\alpha}'''-\frac{1+2m^2}{r^2}\psi_{m,\alpha}''\\
	\label{eq:eigenF_R}
	+&\frac{1+2m^2}{r^3}\psi_{m,\alpha}'+\frac{m^4-4m^2}{r^4}\psi_{m,\alpha} = \omega_{m,\alpha}^4\psi_{m,\alpha}.
\end{align}
With the help of Mathematica \cite{Mathematica}, using a similar procedure to before, we found that
\begin{align}
\notag
	&\omega_{m,\alpha}^4\int_0^a\left(r\psi'_{m,\alpha}\psi'_{m,\beta}+\frac{m^2}{r}\psi_{m,\alpha}\psi_{m,\beta}\right)dr\\
	\notag
	=&\omega_{m,\alpha}^4\int_0^a\left(\frac{m^2}{r}\psi_{m,\alpha}-\psi'_{m,\alpha}-r\psi''_{m,\alpha}\right)\psi_{m,\beta}dr\\
	\notag
	=&\omega_{m,\beta}^4\int_0^a\left(\frac{m^2}{r}\psi_{m,\beta}-\psi'_{m,\beta}-r\psi''_{m,\beta}\right)\psi_{m,\alpha}dr\\
	=&\omega_{m,\beta}^4\int_0^a\left(r\psi'_{m,\alpha}\psi'_{m,\beta}+\frac{m^2}{r}\psi_{m,\alpha}\psi_{m,\beta}\right)dr.
\end{align}
If $\omega_{m,\alpha}\neq\omega_{m,\beta}$ this implies that 
\begin{align}
	\int_0^a\left(r\psi'_{m,\alpha}\psi'_{m,\beta}+\frac{m^2}{r}\psi_{m,\alpha}\psi_{m,\beta}\right)dr=K_{m,\alpha}\delta_{\alpha\beta}
\end{align}
where
\begin{align}
\notag
    K_{m,\alpha} &=\frac{1}{2}\omega_{m,\alpha}^2a^2\Bigg(I_{m-1}(\omega_{m,\alpha}a)I_{m+1}(\omega_{m,\alpha}a)\frac{J_m^2(\omega_{m,\alpha}a)}{I_m^2(\omega_{m,\alpha}a)}\\
    &-J_{m-1}(\omega_{m,\alpha}a)J_{m+1}(\omega_{m,\alpha}a)\Bigg).
\end{align}
Considering the asymptotic expansion of the Bessel function of the first kind Eq.~(\ref{eq:appBesselexp}) and the asymptotic expansion of the modified Bessel function of the first kind, as $x\to\infty$,
\begin{equation}
    I_n(x) = \exp(x)\sqrt{\frac{1}{2\pi x}}\left(1+\frac{1-4n^2}{8x}+\mathcal{O}(x^{-2})\right),
\end{equation}
one can easily see that for large $n$, $K_{n,\alpha}$ increases with $n$ linearly.

In this subsection we (i) calculated the additional surface energy associated with the perturbations used in Eq.~(\ref{eq:3d_pinned_energy}) and (ii) showed that perturbed surface area $K_{n,\alpha}$ increases linearly with $n$, which leads to the use cut-off length scale for wave modes.

% \bibliography{ref}% Produces the bibliography via BibTeX.
\bibliographystyle{apsrev4-2}
\bibliography{zotero}
\end{document}